\documentclass[fleqn,usenatbib]{mnras}

\usepackage{mathptmx}

\usepackage[T1]{fontenc}

\DeclareRobustCommand{\VAN}[3]{#2}
\let\VANthebibliography\thebibliography
\def\thebibliography{\DeclareRobustCommand{\VAN}[3]{##3}\VANthebibliography}

\usepackage{graphicx}	
\usepackage{amsmath}	
\usepackage{amssymb}	
\usepackage{mathtools}
\usepackage{amsfonts}
\usepackage[capitalize]{cleveref}

\title[Cloud microphysics in WASP-39b's atmosphere]{Coupling haze and cloud microphysics in WASP-39b's atmosphere based on JWST observations}

\author[A. Arfaux and P. Lavvas]{
Anthony Arfaux,$^{1}$\thanks{E-mail: anthony.arfaux@univ-reims.fr}
Panayotis Lavvas,$^{1}$
\\
$^{1}$Groupe de spectrom\'erie mol\'eculaire et atmosph\'erique, Universit\'e de Reims Champagne Ardenne, Reims, France\\
}

\date{Accepted XXX. Received YYY; in original form ZZZ}

\pubyear{2023}

\begin{document}
\label{firstpage}
\pagerange{\pageref{firstpage}--\pageref{lastpage}}
\maketitle

\begin{abstract}
We present a study on the coupling of haze and clouds in the atmosphere of WASP-39b. 
We developed a cloud microphysics model simulating the formation of Na$_2$S and MgSiO$_3$ condensates over photochemical hazes in gas giant atmospheres. 
We apply this model to WASP-39b, recently observed with the JWST to study how these heterogeneous components may affect the transit spectrum. 
We simulate both morning and evening terminators independently and average their transit spectra. 
While MgSiO$_3$ formation has negligible impact on the spectrum, Na$_2$S condensates produce gray opacities in the water band, in agreement with HST and JWST observations. 
Moreover, the formation of Na$_2$S on the morning side depletes the atmosphere of its sodium content, decreasing the strength of the Na line. 
Combining morning and evening profiles results in a good fit of the Na observations. 
These nominal results assume a small Na$_2$S/haze contact angle (5.7°).
Using a larger value (61°) reduces the cloud density and opacity, but the effect on the Na profile and spectral line remains identical. 
In addition, the presence of haze in the upper atmosphere reproduces the UV-visible slope observed in the HST and VLT data and contributes to the opacity between the water bands at wavelengths below 2 microns. 
The averaged spectra are rather insensitive to the variation of eddy diffusion and haze mass flux tested in this study, though the UV-visible slope, probing the haze layer above the clouds, is affected.
Finally, our disequilibrium chemistry model, including photochemistry, reproduces the SO$_2$ and CO$_2$ absorption features observed. 
\end{abstract}




\section{Introduction}
\label{Introduction}

Clouds and hazes are expected to form in many exoplanet atmospheres \citep{Sing16,Barstow17,Arfaux22} and strongly affect their surrounding environment \citep{Lavvas21,Steinrueck21,Lee16,Arfaux22,Komacek22a}.
Therefore, in order to understand planetary atmospheres, the comprehension of their formation mechanisms and physical properties, as well as, their impact on the atmosphere, are of prime importance to derive reliable data on the atmospheric structure and composition.
In the solar system, studies have considered the coupling between haze and clouds \citep[e.g.][in Titan's atmosphere]{Lavvas11c}. 
However, most studies on the microphysics of such atmospheric aerosols in exoplanet atmospheres have focused on either haze \citep{Lavvas17,Kawashima19,Adams19,Ohno20,Arfaux22} or clouds \citep{Woitke03,Helling06,Lee15,Powell18,Gao18,Gao20,Gao21,Carone23}, while both are expected to be present and may interact.
Parametric studies can account for both types of opacity \citep{Sing16,Barstow17}, but use \it ad hoc \rm opacity values that are not necessarily representative of the optical and physical properties of haze and clouds, while they neglect the effects related to the complex size and density distributions of the particles.
In this work, we couple cloud and haze microphysics for the first time in the framework of hot-Jupiter atmospheres.
The development of such a coupled description is motivated by the current and forthcoming JWST observations that provide more detailed and precise constraints for the characterization of exoplanet atmospheres.
We investigate as case study the hot-Jupiter WASP-39b that was recently observed with JWST and for which the transit observations suggest that both clouds and hazes may be present in its atmosphere \citep{Arfaux23}.


WASP-39b is a hot-Jupiter exoplanet discovered by \cite{Faedi11} via transit as part of the SuperWASP program.
The large radius and relatively low mass of WASP-39b result in a large scale height, making this planet highly suitable for transit spectroscopy \citep{Fischer16}.
In addition, the hosting star WASP-39A is an 9 Gyr G8 type, therefore suggesting a very weak activity.
The transit spectra obtained for this system are expected to be free from stellar variability effects \citep{Faedi11,Sing16,Fischer16,Ahrer23b,Rustamkulov23}.
We however note that \cite{Pinhas18} did find evidence for imprints of stellar inhomogeneities in the transit observations of WASP-39b, with a covering fraction of $\sim$10\% dominated by cool spots.
They however highlight that stellar contamination is not fully understood and the models might be incomplete or incorrect.
As a consequence of this wide scale height and low stellar contamination, this planet has been widely observed and transit spectra are available from both space-born \citep{Sing16,Fischer16,Wakeford18,Ahrer23a,Ahrer23b,Alderson23,Feinstein23,Rustamkulov23} and ground-based \citep{Ricci15,Nikolov16,Kirk19} facilities covering the spectrum from 0.3 to 5.5 µm (\cref{Tab:W39Obs}).

The first spectroscopic transit observations were obtained by \cite{Ricci15} using U, R and I band filters mounted on the San Pedro M\'artir Telescopes.
These were broad band observations providing little information on the atmospheric structure and composition, but they suggested the absence of extra-atmospheric features (tail, rings, etc.,).
Additional ground-based measurements were provided by \cite{Nikolov16} with the FOcal Reducer and Spectrograph (FORS2) instrument mounted on the Very Large Telescope (VLT), covering the visible range.
Simultaneously, \cite{Sing16} conducted a comparative study of the atmosphere of ten hot-Jupiters among which was WASP-39b.
They provided transit measurements with the Space Telescope Imaging Spectrograph (STIS) on board the HST as well as with the InfraRed Array Camera (IRAC) on board the Spitzer Space Telescope (SST).
Both ground-based \citep{Nikolov16} and space-born observations \citep{Sing16} undoubtedly detected the sodium and potassium lines.
The HST observations were reanalyzed afterwards by \cite{Fischer16}, who found a more shallow UV-visible slope but agreed on the presence of both alkali lines.
Latter (re-)analysis of these observations confirmed the clear detection of these alkali elements \citep{Wakeford18,Fisher19,Pinhas19}.
Although sodium and potassium are detected, the retrieved amounts yield discrepant results with Na mixing ratio ranging from 10$^{-3.86}$ to 10$^{-6.77}$ and K mixing ratio ranging from 10$^{-4.22}$ to 10$^{-7.64}$. 
For both alkali elements, the retrieved values span 3 orders of magnitude from slightly sub-solar ($\times\sim$0.1) to slightly super-solar ($\times\sim$10) with a preference among the different studies for relatively low abundances.
\cite{Wakeford18} conducted observations with the Wide Field Camera 3 (WFC3) on board the HST to extend the observations to the near infrared.
These first observations of the water bands of WASP-39b revealed strong, though slightly muted water features.
The detection of H$_2$O was confirmed by later reanalysis of the HST/WFC3 observations \citep{Tsiaras18,Fisher18,Pinhas19,Min20}, though with discrepant mixing ratios ranging from 10$^{-1.85}$ to 10$^{-5.94}$.
The latest ground-based observations were led by \cite{Kirk19} with the ACAM instrument on the William Herschel Telescope (WHT) and provide transit depths roughly consistent with the previous observations.

The recent survey of WASP-39b with the JWST as part of the Early Release Science (ERS) program, obtained observations with multiple instruments over a wide wavelength range from 0.6 to 5.5 µm.
Observations were conducted with the Near InfraRed Camera \citep[NIRCam,][]{Ahrer23b}, the Near InfraRed Imager \& Slitless Spectrograph \citep[NIRISS,][]{Feinstein23} and the Near InfraRed Spectrograph (NIRSpec) in two different modes: G395H \citep{Alderson23} and PRISM \citep{Ahrer23a,Rustamkulov23}.
The \cite{Feinstein23} observations are in agreement in the visible and near infrared with both ground-based and previous space-born observations while they closely match the \cite{Ahrer23b} observations in the 2.4 - 2.8 µm range.
The \cite{Alderson23} observations are in good agreement with \cite{Ahrer23a} but do not match \cite{Ahrer23b} around 2.8 µm.
We note, overall, that the \cite{Ahrer23b} observations present lower transit depths relative to the other JWST observations in the same wavelength range.
For the NIRSpec PRISM observations, saturation of the detector is observed in the 0.7 - 2.3 µm range, therefore \cite{Ahrer23a} decided to focus on the longer wavelength part of the spectrum (beyond 3 µm), while \cite{Rustamkulov23} managed to work around this issue with a custom bias correction.
\cite{Rustamkulov23} did not detect the potassium line but highlight the saturation of the detector as a possible cause for this non-detection.
 \cite{Ahrer23b,Alderson23} and \cite{Rustamkulov23} observations conclude to the presence of water in the atmosphere of WASP-39b.
Oxidized carbon (CO and CO$_2$) is also detected via these JWST observations, near 4.3 µm, \citep{Ahrer23a,Alderson23,Feinstein23,Rustamkulov23} and confirmed in reanalysis \citep{Tsai23,Carone23,Grant23}.
On the other hand, methane remains undetected and upper bounds on CH$_4$ abundance have been set by \cite{Ahrer23b} and \cite{Rustamkulov23} with limiting mixing ratio of 10$^{-4.26}$ and 10$^{-5.3}$, respectively.
Finally, a feature near 4.05 µm \citep{Ahrer23a,Alderson23} is attributed to SO$_2$, though its thermochemical equilibrium abundance is much smaller than the 10$^{-6}$ to 10$^{-5}$ mixing ratios required to fit the observed feature \citep{Alderson23,Rustamkulov23}.
\cite{Tsai23} used a photochemistry model, assuming a 10$\times$solar metallicity atmosphere, and demonstrated that SO$_2$ can form via photochemical processes in significant amount, able to reproduce the observed feature.
This represents the first direct hint for photochemical processes taking place in exoplanet atmospheres.



\begin{table*}
\caption{Summary of the different transit spectra observations, the facilities with which they were conducted and their wavelength coverage.}
\label{Tab:W39Obs}
\begin{tabular}{c|cc}
Study			&			Facility				&		Wavelength coverage (µm)  	\\
\hline
\cite{Ricci15}		& 	San Pedro M\'artir Telescopes		&		U, R and I bands			\\
\cite{Sing16}		&	 HST STIS \& Spitzer IRAC		&		0.3 - 1 \& 3.6 - 4.5			\\
\cite{Fischer16}		&	 HST STIS \& Spitzer IRAC		&		0.3 - 1 \& 3.6 - 4.5			\\
\cite{Nikolov16}		&			VLT FORS2			&			0.4 - 0.8				\\
\cite{Wakeford18}	&			HST WFC3			&			0.8 - 1.7				\\
\cite{Kirk19}		& 	William Herschel Telescope ACAM	&			0.4 - 0.9				\\
\cite{Ahrer23a}		&		JWST NIRSpec PRISM		&			3 - 5.5				\\
\cite{Ahrer23b}		&			JWST NIRCam			&			2.4 - 4				\\
\cite{Alderson23}	&		JWST NIRSpec G395H		&			2.75 - 5.2				\\
\cite{Feinstein23}	&		JWST NIRISS SOSS		&			0.6 - 2.7				\\
\cite{Rustamkulov23}&		JWST NIRSpec PRISM		&			0.5 - 5.5				\\
\end{tabular}
\end{table*}


%


The recent JWST observations have set better constraints over the widely discrepant results found in previous studies on the metallicity of WASP-39b's atmosphere.
While the considered range of metallicities was spanning from slightly sub-solar \citep[-1 dex,][]{Fischer16} to strongly super-solar \citep[2.45 dex,][]{Kirk19} values, the JWST observations fall in agreement with a slightly super-solar metallicity \citep[0.5 to 1.38 dex,][]{Ahrer23a,Ahrer23b,Alderson23,Feinstein23,Tsai23,Grant23}.
These results are mostly suggested by the absence of CH$_4$ and the strong CO$_2$ and SO$_2$ features detected in WASP-39b's atmosphere indicative of high metallicity \citep{Rustamkulov23,Ahrer23a,Ahrer23b,Tsai23}.
Nonetheless, we note that this super-solar metallicity is inconsistent with the weak sodium and potassium abundances observed, therefore indicating these species must be depleted somehow.
%
%
%
%

The metallicity is only a poor indicator of the relative abundances of the different elements since these can vary depending on the planet formation history \citep{Madhusudhan14b,Fortney20}.
To overcome this issue, we may rely on the abundance ratio of the different elements as the carbon to oxygen ratio.
These two elements are the most abundant (after hydrogen and helium) and the C/O ratio has major ramifications on the atmospheric composition \citep{Molliere15}.
For WASP-39b, the absence of CH$_4$ discussed above indicates a low C/O ratio, which is in agreement with the solar to slightly sub-solar C/O ratio retrieved by most studies \citep{Wakeford18,Kawashima21,Ahrer23a,Ahrer23b,Alderson23,Feinstein23,Tsai23,Crossfield23,Grant23}, with values ranging from 0.2 to 0.55.
\cite{Rustamkulov23} derived an upper limit for the C/O ratio at 0.7, based on a 10$\times$solar metallicity, above which methane would dominate the spectrum beyond $\lambda$ = 1.5 µm.
\cite{Grant23} developed a method for the detection of CO, the main carbon bearing species in hot-Jupiter atmospheres \citep{Woitke18,Fortney20,Arfaux23,Grant23}.
These constraints on the CO, as well as those on water abundances, allow a confident retrieval of the C/O ratio, specifically confirming a solar to sub-solar C/O ratio.
For the other species, elements abundance ratios are not well constrained in WASP-39b's atmosphere.
We only note the super-solar K/O ratio suggested by \cite{Feinstein23}.

Although the first transit observations indicated the absence of haze or clouds \citep{Sing16,Fischer16}, latter reanalysis \citep{Barstow17} as well as additional observations with HST \citep{Wakeford18,Pinhas18,Tsiaras18,Fisher18,Fisher19,Pinhas19} and JWST \citep{Ahrer23a,Ahrer23b,Alderson23,Feinstein23,Rustamkulov23,Carone23} concluded to the presence of haze and/or clouds.
The cold temperature of the planet, relative to other hot-Jupiters, is suited for the formation of cloud species like MnS, Na$_2$S or silicate species \citep{Ahrer23a,Alderson23,Feinstein23}.
\cite{Carone23} conducted a thorough study of cloud composition in WASP-39b's atmosphere with a detailed cloud microphysics model including 16 plausible condensates, coupled to a 3D GCM simulation.
They found a complex cloud composition varying with altitude, as well as, between the morning and evening terminators.
Based on their results, we may expect silicates (MgSiO$_3$, Mg$_2$SiO$_4$ and Fe$_2$SiO$_4$) and metal oxides (SiO, SiO$_2$ and MgO) to dominate the cloud composition in the observed atmosphere, with silicate dominating at the evening terminator and a more balanced composition between these two type of clouds at the morning terminator.
Deeper in the atmosphere, high temperature condensates (TiO$_2$, Fe, FeS, Al$_2$O$_3$ and CaTiO$_3$) are expected to dominate the cloud composition.
These results are obtained assuming heterogeneous nucleation over condensation nuclei of TiO$_2$ and SiO, which are formed via homogeneous nucleation, and do not include haze in the simulation.

The presence of haze is not as consensual with studies preferring grey opacities \citep{Tsiaras18,Fisher18,Ahrer23a,Rustamkulov23,Carone23}, while others require the addition of haze \citep{Barstow17,Pinhas18,Pinhas19,Ahrer23b,Alderson23,Feinstein23}.
Despite these disagreements, the decreasing slope with increasing wavelength observed in the UV-visible range suggests the presence of high altitude absorbers like photochemical haze \citep{Barstow17}.
Although we found no need for haze opacity to fit the observed transit spectrum of WASP-39b assuming a solar metallicity \citep{Arfaux22}, the recent finding of slightly super-solar metallicity for this planet changes this result.
Indeed, with a higher metallicity, the UV slope becomes more shallow, therefore requiring a haze abundance to obtain a decent fit.
In \cite{Arfaux23}, we presented preliminary results, using a 10$\times$solar metallicity, on how the inclusion of Na$_2$S cloud opacity along with hazes may provide a good fit of the recent JWST observations.
These results indicate that, while clouds are necessary to fit the IR region, hazes are still required to fit the UV region and we derived a haze mass flux of 10$^{-15} g.cm^{-2}.s^{-1}$.


In this work, we aim to study the formation of clouds and their interaction with haze particles serving as nucleation site.
For this purpose, we developed a cloud microphysics model, coupled to our self-consistent 1D model \citep{Arfaux22,Arfaux23}, and simulate the formation of Na$_2$S and MgSiO$_3$ condensates over photochemical hazes.
We also account for the effects of cloud formation on the chemistry.
Especially, the formation of sodium sulphide clouds is expected to deplete the atmosphere from its sodium content and thus affect the signature of this species on the transit spectrum, but we also explore how cloud formation affect the abundances of other species partaking in the condensation such as H$_2$O and H$_2$S.
We conduct our simulations for both morning and evening terminators independently to study how the temperature differences of the two terminators may affect the cloud formation and the chemical composition, as well as the resulting transit spectrum.

Our model is described in \cref{Sec:Method} with \cref{Sec:CldCompo} discussing the clouds microphysical properties, as well as, the cloud microphysics and prototype models.
\cref{Sec:Spectrum} discusses the theoretical transit spectra calculation and the cloud optical properties, as well as, the combination of morning and evening terminator results.
The results for the best fit case are detailed in \cref{Sec:Results}, with \cref{Sec:HazeClouds} focusing on the haze and cloud coupling and \cref{Sec:Chemistry} detailing the chemical composition.
Sensitivity tests are conducted and summarized in \cref{Sec:Sensitivity}, with \cref{Sec:STCA} exploring the effect of the surface tension for MgSiO$_3$ formation and the contact angle for Na$_2$S, while \cref{Sec:Eddy} and \cref{Sec:HzMF} study the effect of changing the eddy diffusion and haze mass flux, respectively.
We discuss our results in \cref{Sec:Discussion} and outline the main conclusions in \cref{Sec:Conclusions}.

\section{Method}
\label{Sec:Method}

\begin{figure}
\includegraphics[width=.5\textwidth]{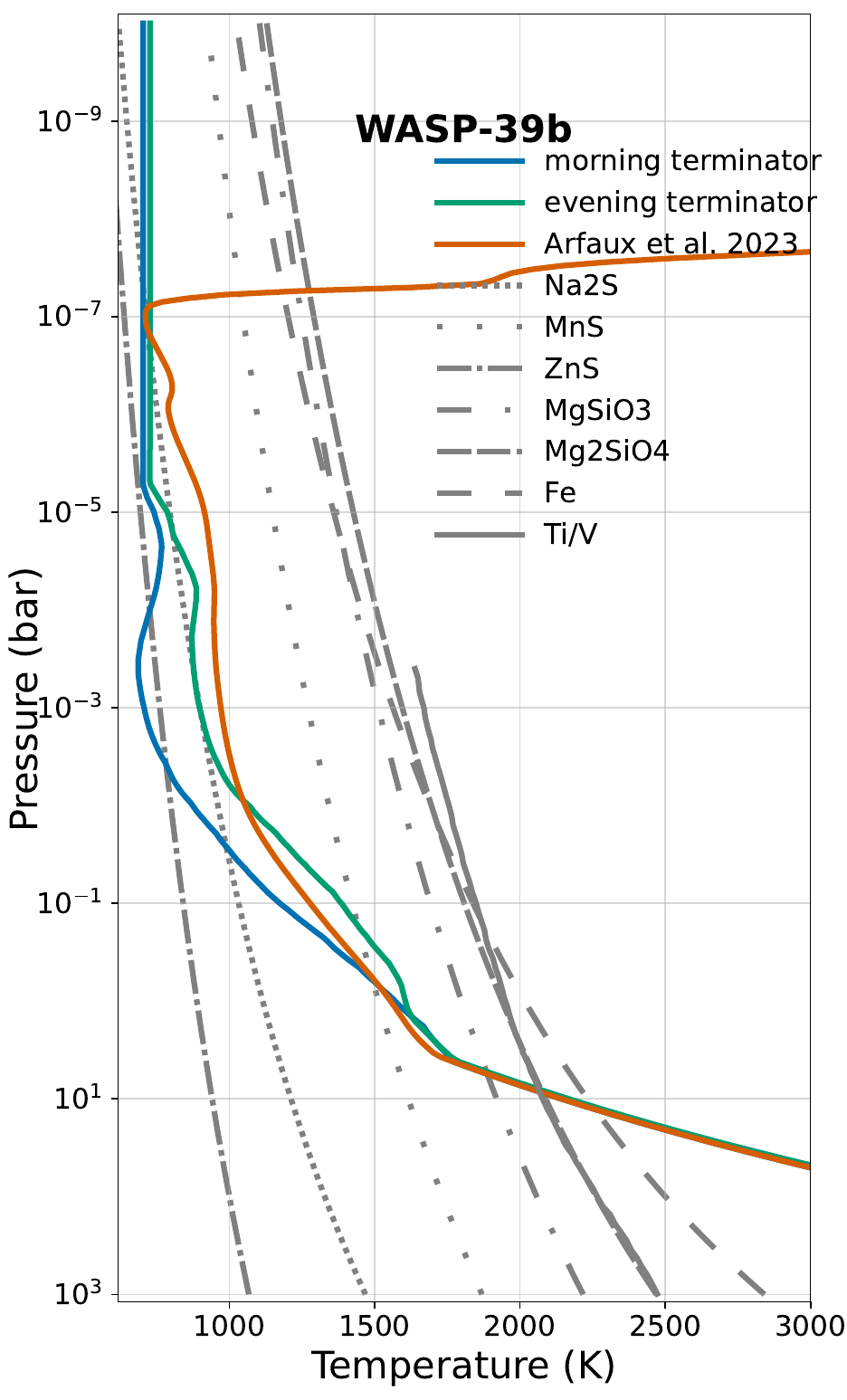}
\caption{Temperature profiles used in our study for the morning (blue line) and evening (green line) terminators, as well as the profile derived for WASP-39b in \protect\cite{Arfaux23} (orange line). 
The morning and evening terminator profiles are taken from \protect\cite{Tsai23}, while the orange profile was calculated assuming a full heat redistribution with a 1D radiative/convective model. 
Overplotted in gray lines are condensation curves for cloud condensate candidates \protect\citep[][]{Visscher06,Visscher10,Fortney08}.}
\label{Fig:pT}
\end{figure}

\subsection{Expected cloud composition}
\label{Sec:CldCompo}

Many different condensates can form in planetary atmospheres depending on the pressure-temperature conditions and bulk composition \citep[][GGchem thermochemistry model]{Woitke18}.
In our work, we use the p-T profile obtained by \cite{Tsai23} for evening and morning terminators based on GCM simulations.
Those are roughly the same in the deep atmosphere below the 1 bar altitude (\cref{Fig:pT}).
The main differences occur between 1 bar and 10 mbar with a $\sim$200 K hotter atmosphere on the evening terminator.
Above the 10 mbar altitude, the two temperature profiles converge to similar values.
We see that, in both evening and morning terminator conditions, the formation of iron, titanium or silicate clouds is likely to happen in the deep atmosphere with a cloud base at 10 bar.
At higher altitudes, the formation of MnS is expected for both terminators around 0.1 bar, while ZnS and Na$_2$S may form only on the morning side with cloud bases at 4 and 30 mbar, respectively.
We note that the formation of Na$_2$S at high altitude can be expected on the evening terminator, however, the p-T profiles used do not demonstrate the presence of the anticipated thermosphere (orange line in \cref{Fig:pT}) which would prevent the formation of these clouds in that region of the atmosphere.

As previously discussed, the super-solar metallicity derived for WASP-39b would lead to a stronger sodium line than observed with HST and JWST, indicating the need for a loss mechanism for Na.
Disequilibrium chemistry can partially reduce the Na atmospheric abundance but our simulations show that this loss is not sufficient to explain the observed Na transit signature \citep{Arfaux22}.
Given the terminator temperature profiles and according to thermochemical equilibrium calculations conducted with GGchem \citep{Woitke18}, we consider that the loss of sodium to Na$_2$S cloud formation is a reasonable hypothesis and we explore this possibility in the current study.

In this work, we aim to study clouds in the observable part of the atmosphere.
According to the results of \cite{Carone23}, it is therefore unlikely to have the formation of condensates including Ca, Ti, Al or Fe in the observed region of the atmosphere.
We may however expect silicate condensates as MgSiO$_3$ and Mg$_2$SiO$_4$.
Based on preliminary results we obtained with the thermochemical equilibrium model GGchem, including condensation and rainout, we observe that MgSiO$_3$ is more likely to form in the observable region than Mg$_2$SiO$_4$.
In addition, the formation of these silicate clouds is expected to reduce the abundance of water, thus affect the resulting UV-visible slope of the transit spectrum \citep{Lavvas17}.
Indeed, this condensate is formed via the reaction of Mg, SiO and H$_2$O, forming H$_2$ as secondary product \citep{Visscher10}, and we therefore decided to include MgSiO$_3$ in our calculations.
We exclude deep atmosphere clouds from our calculation since those are formed with Fe, Ca, Ti or Al, which present lower abundances relative to Si and therefore, their formation is expected to have a weaker impact on the chemistry and spectra than the formation of MgSiO$_3$. 

The abundance of sodium is much larger than that of manganese or zinc, therefore Na$_2$S would be the dominating sulphur condensate on the morning terminator.
MnS condensates might be expected on the evening terminator, where they would be the main sulfur-containing condensates.
However, though MnS forms higher up compared to MgSiO$_3$, it is 100$\times$ less abundant than Si and is not expected to expand in the observed atmosphere.
Therefore MnS condensates would not affect the morning terminator and are not anticipated to impact the spectra.
We therefore decided not to include this species in our model and we discuss this decision further below.


Our cloud microphysics model derives the cloud distribution accounting the transport of the condensing species and cloud particles, as well as, their nucleation and growth mechanisms.
The clouds studied here, MgSiO$_3$ and Na$_2$S, are non-molecular condensates characterized by the absence of the corresponding condensing species in the gas phase.
For non-molecular condensates, cloud formation and growth happens via surface reactions of chemical products present in the gas phase \citep{Woitke03,Helling06}, thus requiring the presence of condensation nuclei.
\cite{Woitke03} and \cite{Helling06} developed models of cloud formation to study Brown Dwarf atmospheres, simulating chemical reactions happening on the surface of condensation nuclei.
\cite{Lee15} used a similar model applied to the study of cloud formation in the hot-Jupiter exoplanet HD-189733b.
These models account for a first step of homogeneous cloud nucleation, to form the condensing nuclei.
They then consider the growth, that starts by the adsorption of the condensing molecules and their diffusive transport over the surface to form the condensate via surface reactions.
However, in our model, we decided to use a slightly more simple way of accounting for the formation of such clouds using the classical nucleation theory via proxy species \citep{Pruppacher97,Powell18,Chachan19,Gao20}.
Let A and B be the required gas-phase chemical species that react on the surface of the particle to form AB[\,s]\, condensate.
If A is present in much weaker abundance than B ($\chi_A \ll \chi_B$), then the abundance of B will be negligibly affected and the collision rate of species A on the surface of the particle will be the limiting factor for cloud formation and growth.
Therefore, we can use the species A as a proxy for the AB[\,s]\, condensate formation and use the classical nucleation theory assuming A is the only condensing species.
Recently, \cite{Lee23b} developed "mini-cloud", a cloud microphysics model implementing this assumption and designed to be coupled to GCMs, and studied the cloud composition of HAT-P-1b's atmosphere.

For MgSiO$_3$, the limiting species is SiO and the saturation pressure is therefore calculated from the abundances of this species from \cite{Visscher10}:
\begin{equation} 
\begin{aligned}
log (\chi_{SiO}) &= log (\frac{P_{SiO}}{P}) = 13.67 - \frac{28\;817}{T} - log P - [Fe/H]\\
P_{SiO} &= P \times 10^{13.67 - \frac{28\;817}{T} - log P - [Fe/H]}
\end{aligned}
\end{equation} 
with  the metallicity fixed to 10$\times$solar: $[Fe/H] = 1 dex$ and the pressure in bar.
We use a volume density of 4.103$g.cm^{-3}$ for MgSiO$_3$ and a surface tension of 1280 $dyne.cm^{-1}$ \citep[][estimation based on Mg$_2$SiO$_4$]{Powell18}.
Finally, the latent heat is fixed to 1.543$\times 10^{11} erg.g^{-1}$ \citep{Chase98}.

Na$_2$S is formed via the reaction of Na with H$_2$S, the former being less abundant than the latter by a factor of $\sim$10, it is therefore the limiting species and the saturation pressure for the formation of these clouds is taken as that of atomic sodium from \cite{Morley12}:
\begin{equation} 
P_{Na} = 10^{8.55 - \frac{13\;889}{T} - 0.5[Fe/H]}
\end{equation}
with the metallicity fixed at $[Fe/H] = 1 dex$.
As for the silicate clouds, the latent heat is taken from \cite{Chase98} ($L = 4.691\times10^{10} erg.g^{-1}$).
The volume density used is 1.856$g.cm^{-3}$ and the surface tension is approximated at $\sigma$ = 100 $dyne.cm^{-1}$ based on measurements for Na$_2$SO$_4$ \citep{dos-Santos10} and NaCL \citep{Lee18}.

The cloud formation process starts with the nucleation of the condensing species over a seed (heterogeneous nucleation).
The particles of condensing species collide and stick to the nuclei.
We consider that the adsorbed molecules can migrate over the surface resulting in one single wetted area, called an embryo.
Within supersaturated conditions, this embryo will grow and reach a limit radius above which the particle will quickly grow and condensation will take over nucleation.
This limit size is called the germ radius denoted $a_g$.
The ability of the molecules to stick to the surface depends on the interactions between the nuclei material and the condensing species.
This is accounted through the contact angle ($\theta_c$), that is the angle formed by the gas/condensate and the nucleus/condensate interfaces where they join each other at the edge of the embryo.
A small value of the contact angle (between 0° and 90°) indicates a good affinity between the condensing and the nucleus materials while a large value (between 90 and 180°) indicates a bad sticking efficiency.
Therefore, a $\theta_c$ lower than 90° corresponds to a wettable material, while a larger $\theta_c$ corresponds to a non-wettable material.
On the following, we use the cosine of the contact angle, called the wetting coefficient, which can vary from -1 ($\theta_c = \pi$), meaning a very poor affinity, to 1 ($\theta_c = 0$), meaning a very good affinity.

In this work, we consider hazes as the nucleation sites.
However, no lab experiment on the nucleation of MgSiO$_3$ and Na$_2$S over soot-type aerosols has been done so far and the contact angle between the haze particles and the condensing material is unknown.
Most studies on the formation of silicate clouds use very low contact angles \citep[$\theta_c$ = 0.1°]{Powell18,Gao18}.
These low values are supported by \cite{Gao20} and \cite{Gao21} who estimated contact angles smaller than 0.1° for the heterogeneous nucleation of silicate condensate over TiO$_2$ particles, based on their respective surface tensions.
On the other hand, \cite{Gao20} found a contact angle of 61° for the nucleation of Na$_2$S over TiO$_2$ clusters.
In our current calculation, we consider, for both condensate types, the value of m = 0.995 (corresponding to a contact angle of 5.7°) derived by \cite{Lavvas11c} for the wetting coefficient of methane over tholin particles in Titan's atmospheric conditions, as a rough estimate.
This is a conservative approach for MgSiO$_3$ as we use a much larger contact angle compared to the value usually considered for this species formation on TiO$_2$.
However, this may overestimate the nucleation rates for Na$_2$S condensates.
We however highlight that the soot composition for the haze particles considered in this study is likely to behave in a different way than TiO$_2$ and that the interaction of the haze particles with the condensed phase remains unknown.
Our purpose here is to explore how the formation of Na$_2$S clouds may affect the interpretation of transit observations. Thus we treat the Na$_2$S contact angle as a free parameter and explore its impact on the resulting cloud properties and transit spectra.

Once particles are formed via this nucleation process, condensation and evaporation of the particles will drive their size distribution, while transport will spread them away from their formation region, where they will eventually face sub-saturation conditions leading to the loss of the particles.
Our cloud microphysics calculation process is based on \cite{Pruppacher97} and have been applied to the study of condensation in Titan's \citep{Lavvas11c} and Pluto’s \citep{Lavvas21a} atmospheres.
The details of the calculations can be found in these works.

The formation of these condensates affect the chemistry by removing the species involved in the process.
We therefore couple our cloud microphysics model to a self-consistent 1D model that simulates exoplanet atmospheres accounting for disequilibrium chemistry, haze microphysics and radiative-convective energy transfer.
The chemistry model assumes a C/O ratio of 0.457 and a 10$\times$solar metallicity (1 dex) value consistent with the most recents studies based on JWST observations of WASP-39b.
This model allows to study the different feedbacks between the haze, the chemistry and the radiation field.
For example, the presence of haze particles impacts how the light form the host star penetrates the atmosphere, therefore affecting the photochemistry.
With the coupling to the cloud microphysics, we can take into account the removal of the SiO, Na, H$_2$O and H$_2$S.
Our model includes a physically derived parameterization of the eddy profile that accounts for convective mixing in the deep atmosphere and gravity waves in the upper atmosphere \citep{Arfaux23}.
This prototype model is further described in \cite{Lavvas14,Lavvas17,Lavvas21,Arfaux22}.

\subsection{Transit spectrum}
\label{Sec:Spectrum}


The theoretical transit spectra are calculated with a spectral model taking into account various opacity sources.
The list of gaseous opacity sources is provided in \cite{Lavvas21} to which we added SO$_2$ absorption \citep{Underwood16}.
The model also includes Rayleigh scattering by the main atmospheric species as well as Mie scattering and absorption by haze and cloud particles.
We consider a soot composition for the haze particles as they can possibly survive the extreme temperature conditions encountered in hot-Jupiter's atmospheres \citep{Lavvas17}.
The cloud refractive indices are obtained from \cite{Montaner79,Khachai09} for Na$_2$S and \cite{Scott96} for MgSiO$_3$.
Additional details on the transit simulation are provided in \cite{Lavvas17} and \cite{Arfaux22}.

In order to fit the observations, the spectrum is referenced to the \cite{Feinstein23} observations.
In a nutshell, this means that the spectrum is shifted to match the mean value of those observations.

In our process, we simulate both morning and evening terminators separately.
We therefore need to reconstruct the averaged spectrum as both terminators will affect the transit.
The method we chose is to calculate the mean of the morning and evening transit depths.
This assumes a sharp connection of the terminators where the planets is made of two perfect hemispheres of radii $R_{mr}$ and $R_{ev}$.
The apparent surface of the planet is therefore the sum of these two hemispheres:
\begin{equation}
S_p = \frac{\pi R_{mr}^2}{2} + \frac{\pi R_{ev}^2}{2} = \frac{\pi}{2} ( R_{mr}^2 +  R_{ev}^2 )
\end{equation}
The transit depth is the ratio of the flux blocked by the planet ($F_{out} - F_{in}$) to the flux of the star out of transit ($F_{out}$):
\begin{equation}
T_D = \frac{F_{out} - F_{in}}{F_{out}}
\end{equation}
where $F_{out}$ is the out of transit flux, proportional to the surface of the star ($S_* = \pi R_*^2$) and $F_{in}$ is the in transit flux, proportional to the difference between the star and the planet surfaces ($S_* - S_p$).
We therefore obtain:
\begin{equation}
\begin{aligned}
T_D &= \frac{S_* - (S_* - S_p)}{S_*} = \frac{S_p}{S_*} = \frac{\pi}{\pi R_*^2} \frac{( R_{mr}^2 +  R_{ev}^2 )}{2}  \\
&= \frac{1}{2} \left[ \left( \frac{R_{mr}}{R_*}\right)^2 +  \left( \frac{R_{ev}}{R_*}\right)^2 \right] = \frac{1}{2} ( T_D^{mr} + T_D^{ev} )
\end{aligned}
\end{equation}
where $T_D^{mr}$ and $T_D^{ev}$ are, respectively, the transit depths calculated for the morning and evening terminators independently.

\section{Nominal case study}
\label{Sec:Results}

\begin{figure}
\includegraphics[width=0.5\textwidth]{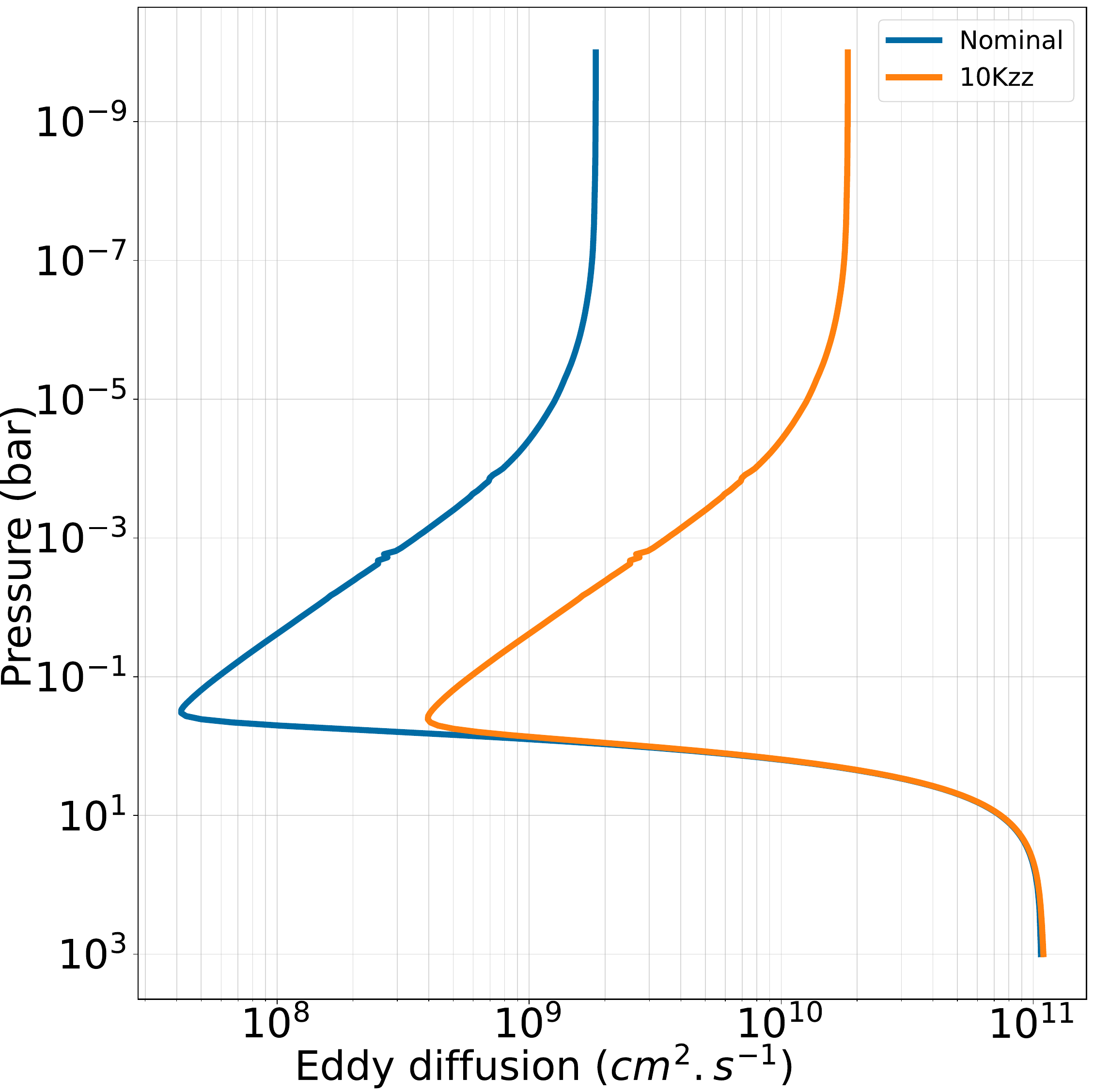}
\caption{Eddy diffusion profiles used in the different cases. }
\label{Fig:Eddy}
\end{figure}

\begin{figure*}
\includegraphics[width=0.49\textwidth]{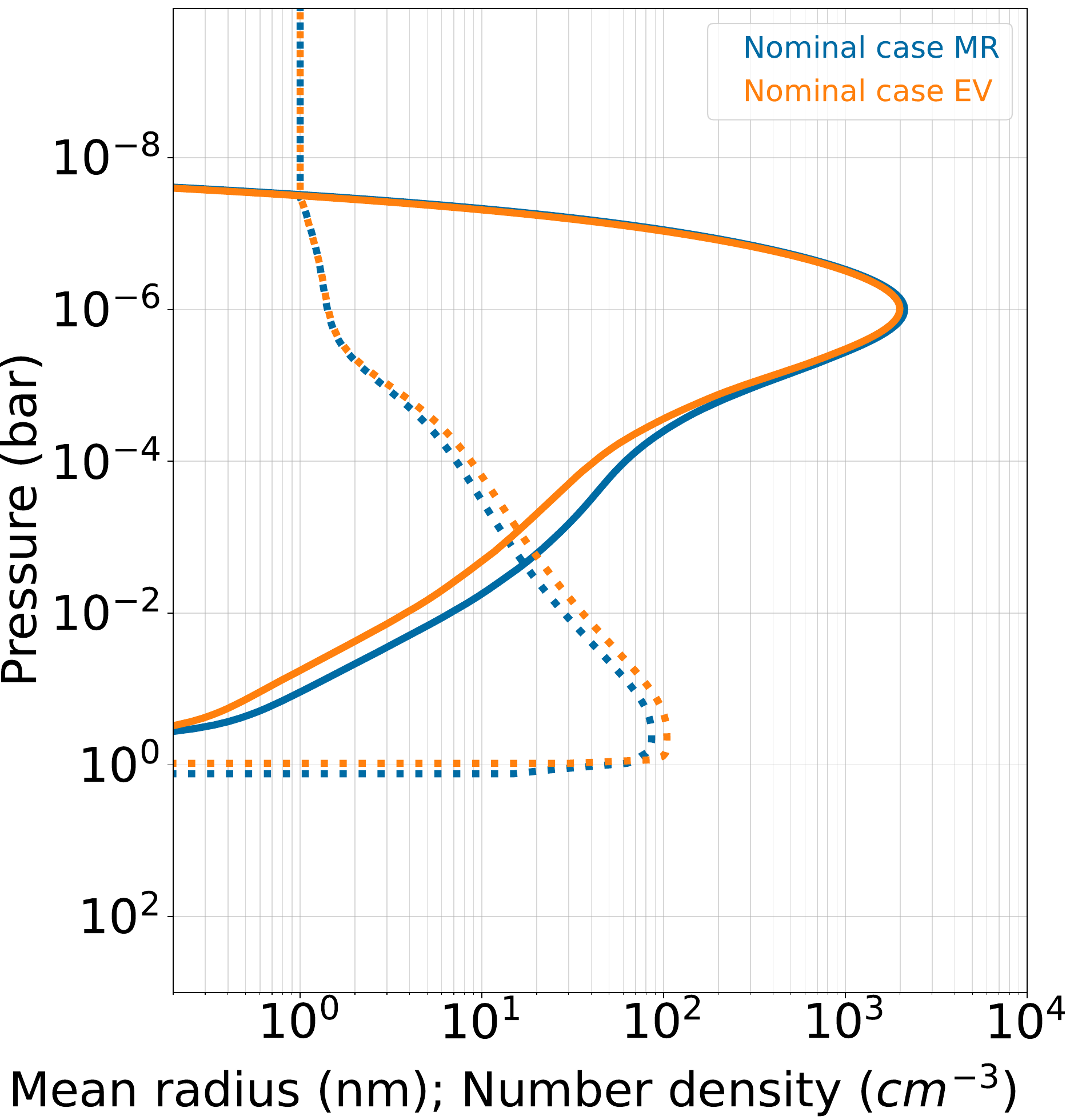}
\includegraphics[width=0.49\textwidth]{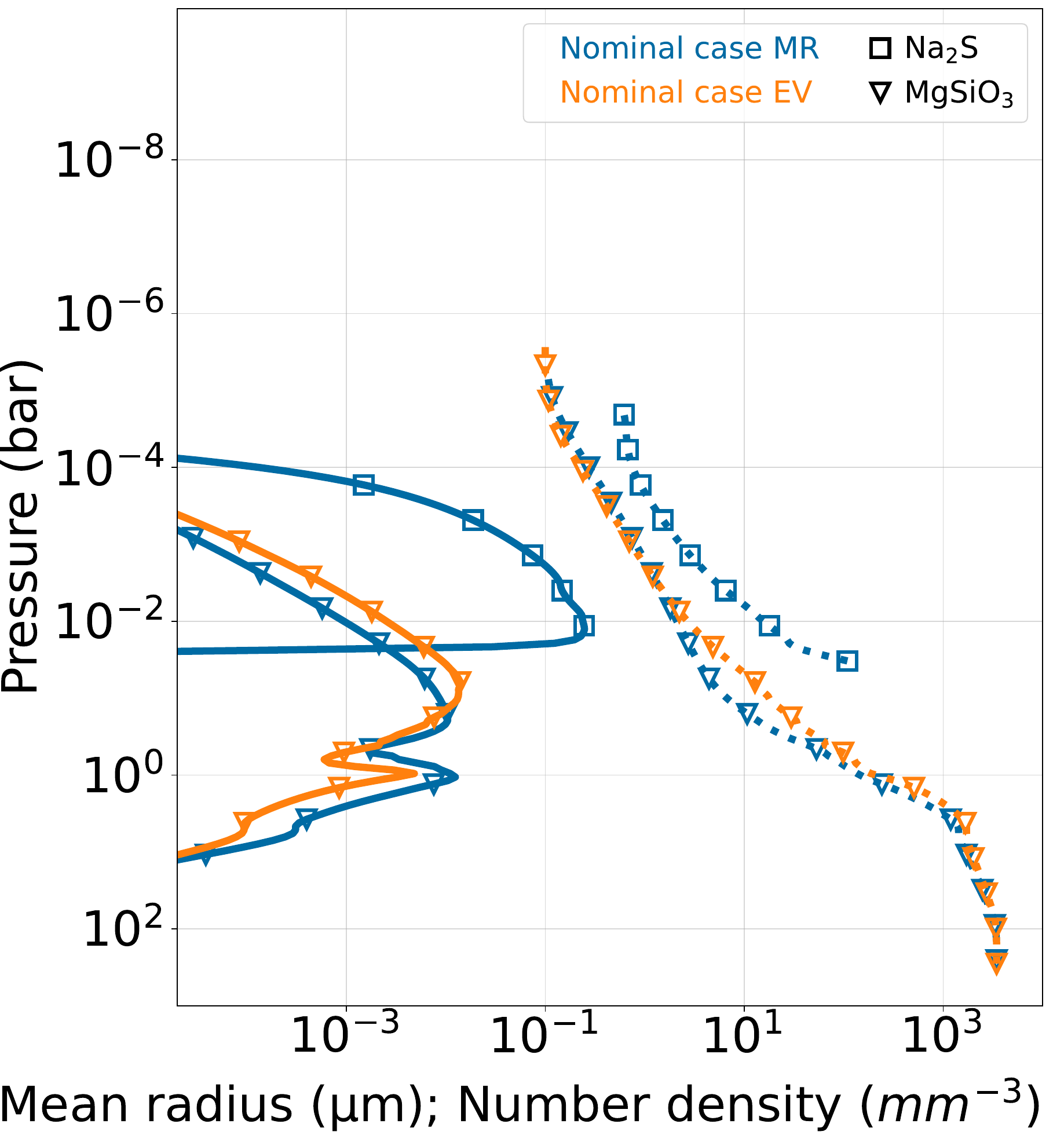}
\includegraphics[width=0.99\textwidth]{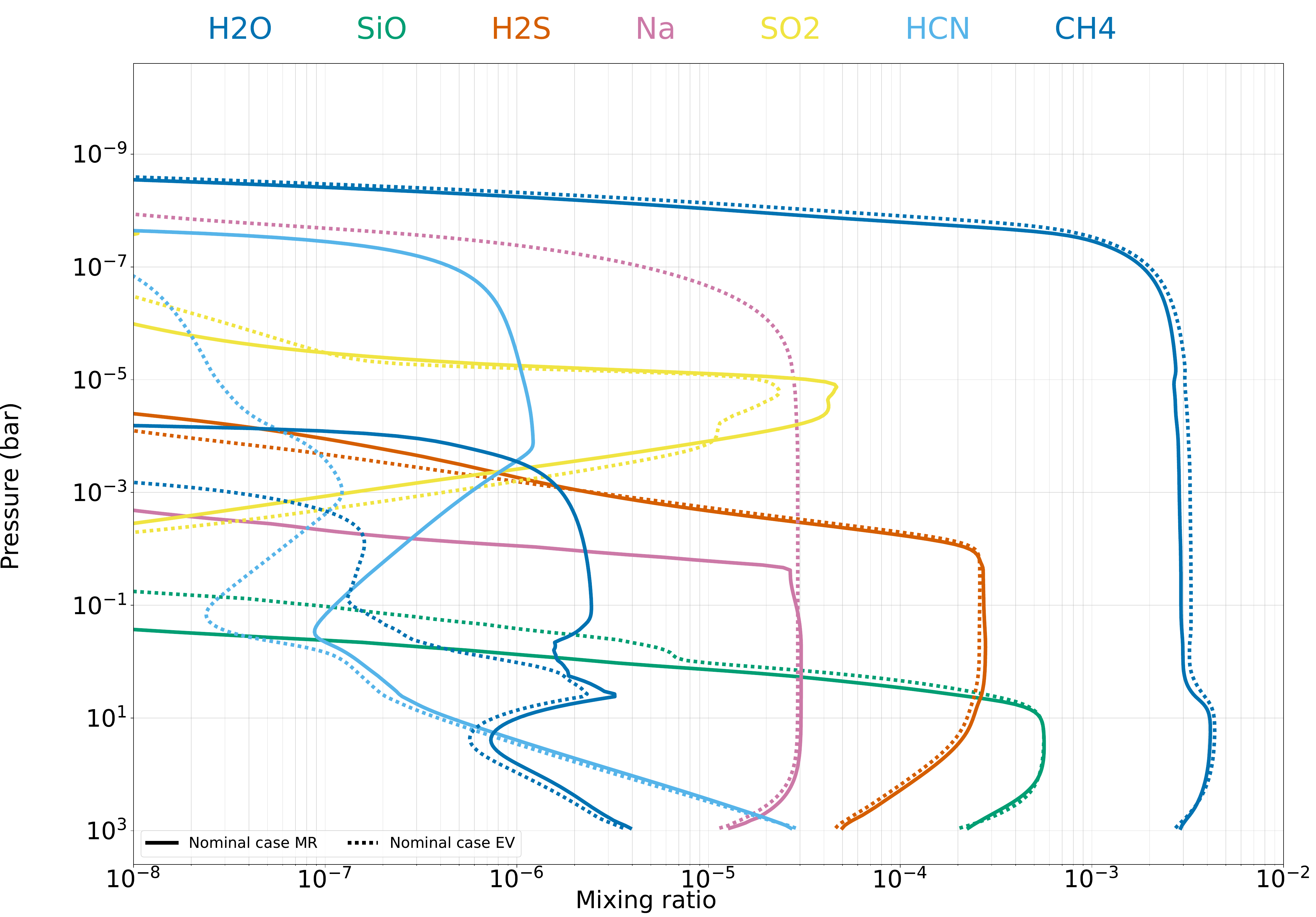}
\caption{
Haze distribution (upper left panel), cloud distribution (upper right panel) and chemical composition (bottom panel) of the best-fit case.
For the haze and cloud distributions, dotted lines are for the mean particle radii and solid lines for the particle densities.
The blue lines are for the morning terminator and the orange lines for the evening.
For the cloud distribution, triangles refer to MgSiO$_3$ condensates and squares to Na$_2$S. 
The solid lines are for the morning terminator and the dotted lines for the evening.
}
\label{Fig:Best}
\end{figure*}

\begin{figure}
\includegraphics[width=0.5\textwidth]{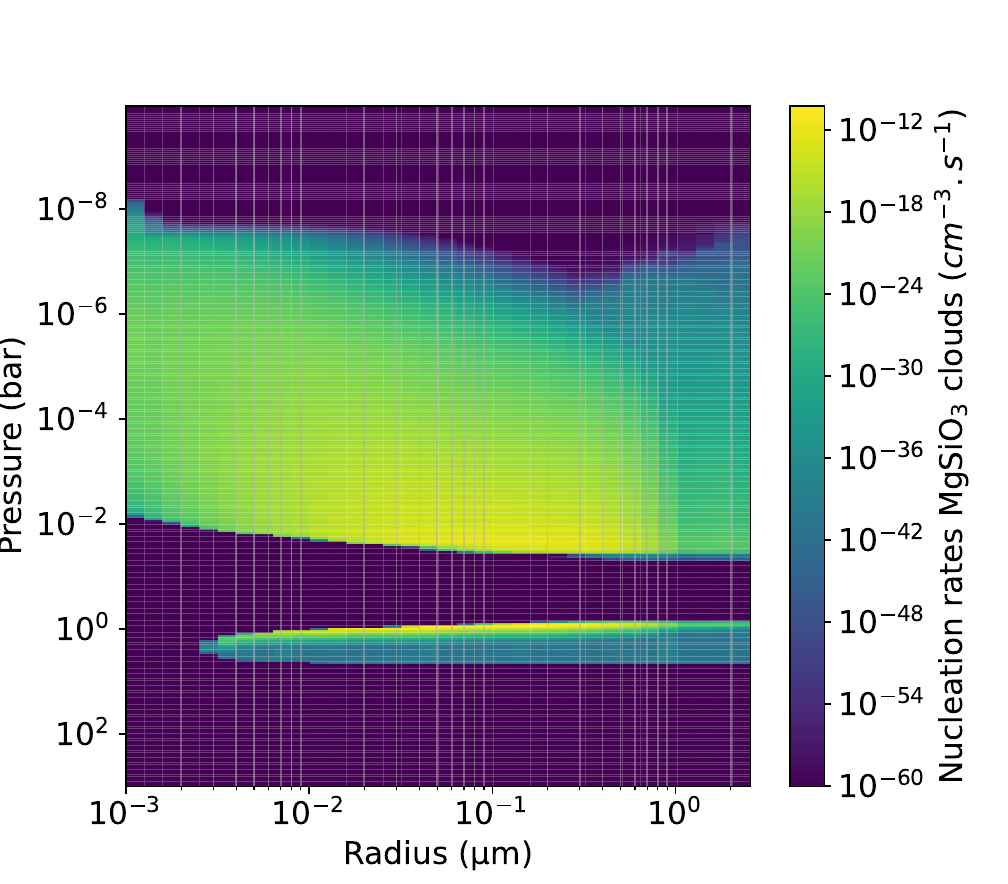}
\includegraphics[width=0.5\textwidth]{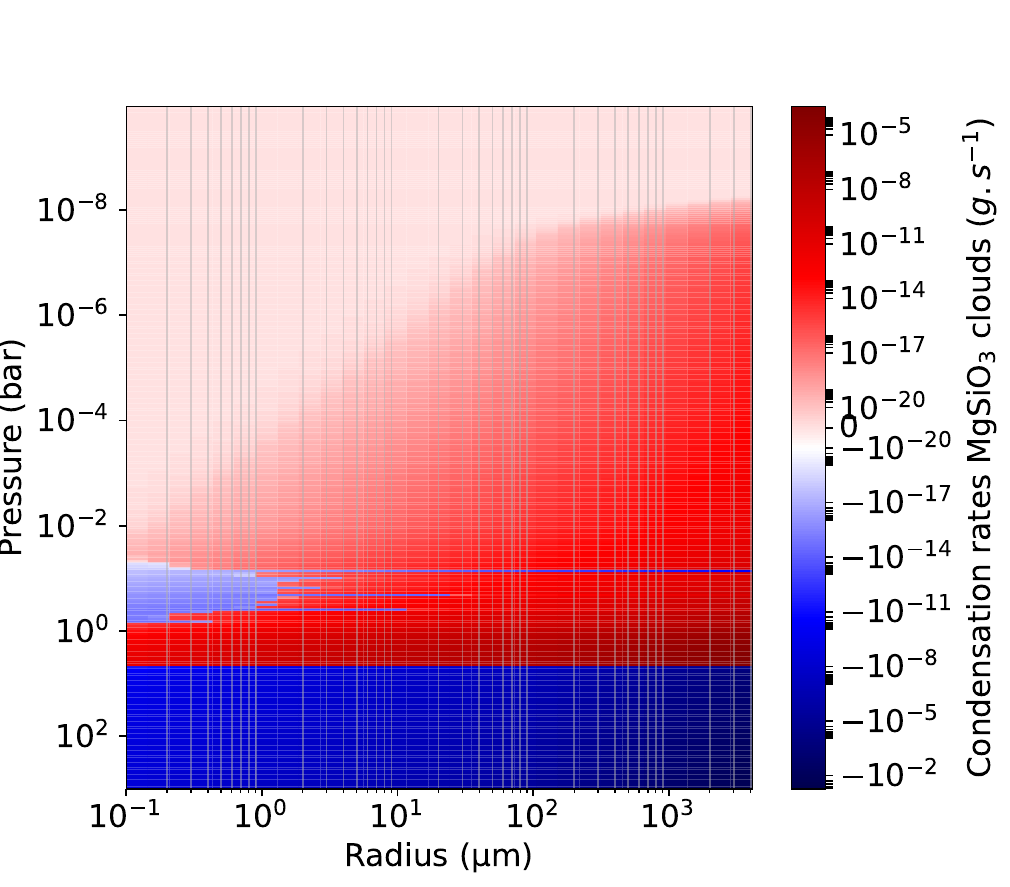}
\caption{Top panel: production rates of MgSiO$_3$ particles due to nucleation for the evening terminator nominal case. The x axis is the nucleus size. Bottom panel: condensation rates of MgSiO$_3$ for the different cloud particle size.}
\label{Fig:NucCondRatesMgSiO3}
\end{figure}

\begin{figure*}
\includegraphics[width=\textwidth]{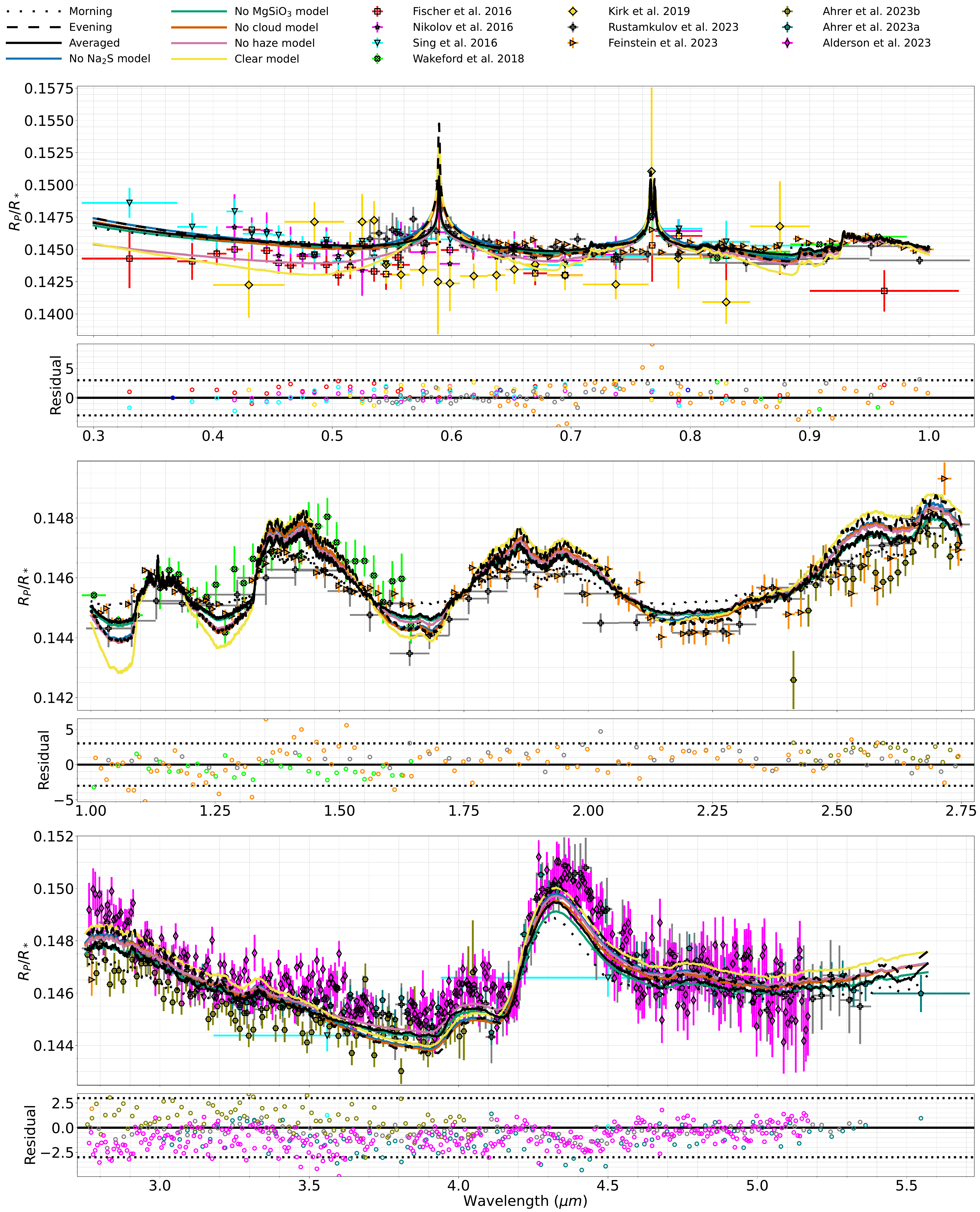}
\caption{Transit spectra for the nominal case (black lines), the morning side spectrum is shown in dotted line, the evening in dashed line and the averaged in solid line. 
The colored lines are additional results for the best-fit conditions, removing some of the heterogenous opacities.
The residuals are for the nominal averaged spectrum.
The simulated transit spectra are smoothed with a savgol filter.}
\label{Fig:SpectraBest}
\end{figure*}

\begin{figure}
\includegraphics[width=0.5\textwidth]{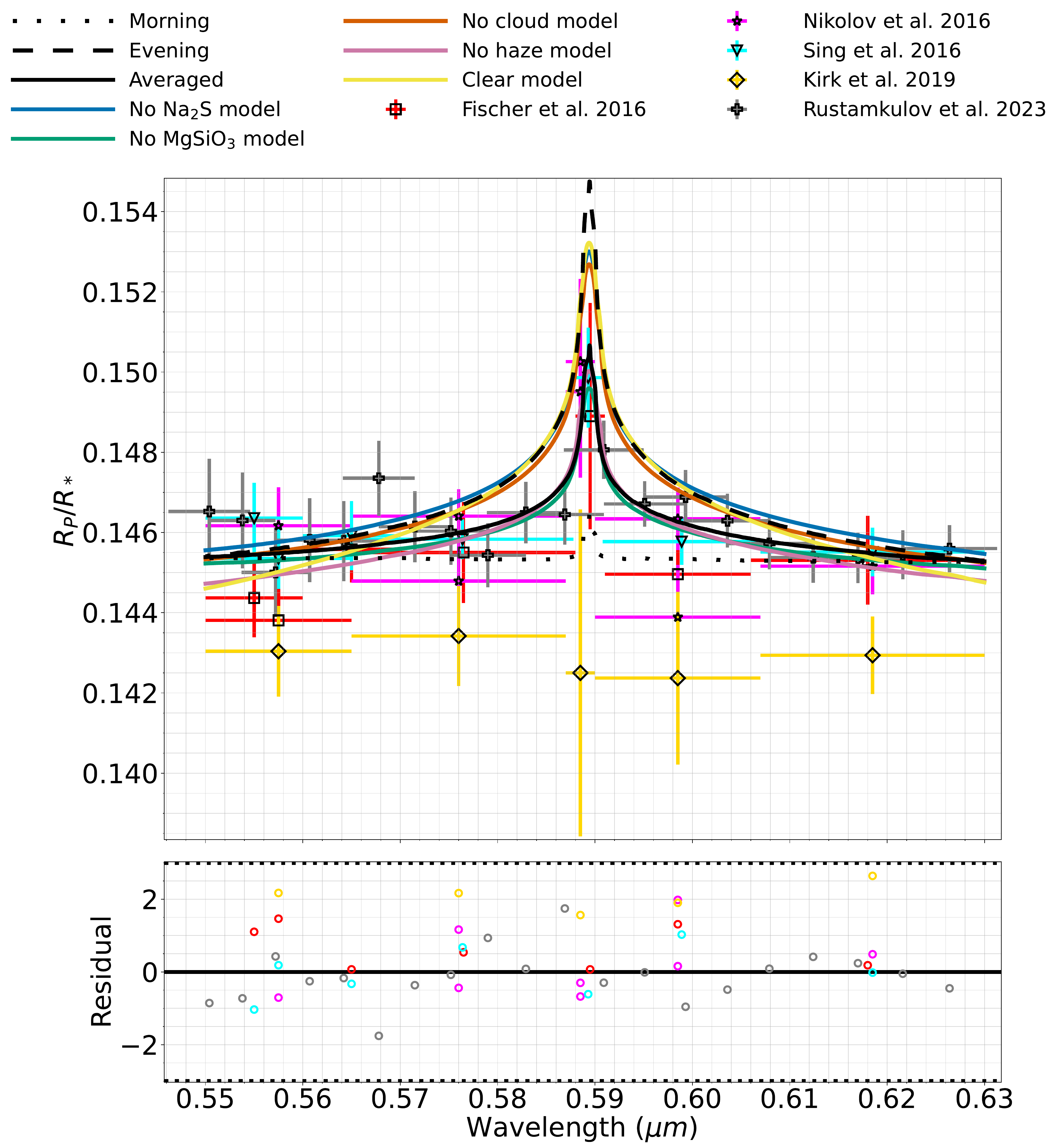}
\caption{Transit spectra around the Na line for the nominal case (black lines), the morning side spectrum is shown in dotted line, the evening in dashed line and the averaged in solid line. 
The colored lines are additional results for the best-fit conditions, removing some of the heterogenous opacities.
The residuals are for the nominal averaged spectrum.
The simulated transit spectra are smoothed with a savgol filter.}
\label{Fig:SpectraBestNa}
\end{figure}

Here we focus on the nominal case, obtained with the eddy profile calculated with our parameterization \citep{Arfaux23}, shown in blue line in \cref{Fig:Eddy} and assuming a haze mass flux of 3$\times$10$^{-15} g.cm^{-2}.s^{-1}$.

\subsection{Haze and clouds}
\label{Sec:HazeClouds}

Hazes form in the upper atmosphere around the 1 µbar altitude and settle down.
The particles are produced with a radius of 1 nm and then coagulate during settling to form larger particles (\cref{Fig:Best}).
We therefore observe a particle number density decreasing from 2,000 cm$^{-3}$ to < 1 cm$^{-3}$ as we move from 1 µbar to 0.1 bar, while the mean particle size increases from 1 nm to 100 nm.
Both terminators demonstrate similar results. 
We only note slightly larger haze particles for the evening side with a slightly lower particle number density.
This is related to the stronger coagulation produced by the hotter temperatures found on the evening terminator.

Cloud formation starts with the nucleation of condensing material over the haze particles, therefore requiring both supersaturation conditions and the presence of large condensing nuclei.
The size of the latter required for the formation of clouds mostly depends on the Kelvin effect, that is the increase of the saturation pressure over a curved surface.
This effect implies that the smaller the nucleation site, the larger the supersaturation conditions required for the nucleation and condensation.
Therefore, nucleation is more efficient as we move deeper in the atmosphere where the haze particles have coagulated into larger sizes (\cref{Fig:Best}).
The highest nucleation rates therefore occur at the cloud base, located near 10 mbar for Na$_2$S and 100 mbar for MgSiO$_3$, and correspond to a peak in the density profile of the clouds.
For Na$_2$S, the undersaturation conditions below 10 mbar result in the sublimation of the particles. 
However, we observe a different behavior for MgSiO$_3$ with the presence of a second, separated, nucleation region deeper in the atmosphere around 1 bar (\cref{Fig:NucCondRatesMgSiO3}).
In addition, between the two nucleation region, we note the sublimation of the smaller cloud particles but positive condensation rates for the larger particles.
Therefore, MgSiO$_3$ particles are still present below the "cloud base" with a second, weaker, peak in the cloud distribution due to the deeper nucleation region.

Comparing the terminators, we observe a small difference in the location of the MgSiO$_3$ peak between the terminators, with slightly lower pressures for the evening case, though a similar particle density (\cref{Fig:Best}).
This relates to a sublimation happening at higher altitude due to the hotter temperatures of the evening case.
However, for the secondary peak at 1 bar, we observe a lower number density for the evening case.
As the temperature is roughly the same at that depth between evening and morning, the difference is related to the less numerous haze particles found in this region for the evening terminator.
For Na$_2$S the hot temperatures of the evening terminator produce saturation pressures larger than the Na partial pressure, preventing the formation of this condensate.

The newly formed particles are mixed to the upper atmosphere where they are observable and down to the region where they sublimate due to undersaturation conditions.
As they move and cross through supersaturated regions, they accrete the condensing species, and grow up to reach mean radii up to a few hundred microns for Na$_2$S and a few thousand microns for MgSiO$_3$.
We note that large particles are more inclined to undergo gravitational settling while smaller particles are more efficiently mixed both downwards and upwards.
As a consequence, we obtain a mean cloud radius decreasing with altitude with a maximum at the cloud base (\cref{Fig:Best}). 
MgSiO$_3$ condensates reach higher altitudes ($\sim$1 µbar) than Na$_2$S ($\sim$10 µbar), though in much lower abundances and are not detectable.
Cloud formation also affects the haze distribution, though the changes are rather negligible, since only a small part of the particles actually serves as nucleation site, and the differences in terms of mean radius and number density are hardly observable.

The transit spectrum for this nominal case is shown in black line in \cref{Fig:SpectraBest} (the dotted line corresponds to the morning terminator spectrum, the dashed line to the evening and the solid line to the averaged), along with additional spectra removing some of the heterogenous opacities, therefore allowing to observe the impact of the haze and clouds.
MgSiO$_3$ clouds become optically thick at higher pressures ($\sim$ 100 mbar) than Na$_2$S ($\sim$ 2 mbar).
We therefore observe that Na$_2$S has a much stronger impact on the spectrum than MgSiO$_3$ (\cref{Fig:SpectraBest}).
Indeed, the model without MgSiO$_3$ condensates (green line in \cref{Fig:SpectraBest}) overlaps with the nominal model (black line), indicating negligible effects from this condensate type.
Na$_2$S, however, strongly affects the spectrum in the water band.
We effectively observe that the model excluding Na$_2$S condensates (blue line) provides weaker transit depths in the gaps between the water bands.
We further note that the Na$_2$S-free model (blue line) and the cloud-free model (orange line) overlap, supporting that Na$_2$S dominates the changes related to cloud opacities, while MgSiO$_3$ condensates have negligible impact on the spectrum.

The absence of Na$_2$S opacity on the evening side provides opposite behavior between the morning and evening terminator spectra in the water bands (\cref{Fig:SpectraBest}).
Since the formation of MgSiO$_3$ has a negligible impact on the spectrum, the evening transit spectrum is close to a cloud-free atmosphere, therefore underestimates the transit depth in the gaps between the water bands, particularly at 1.05 and 1.25 µm (black dashed line in \cref{Fig:SpectraBest}).
On the other hand, the presence of Na$_2$S condensates on the morning terminator results in more muted water absorption compared to the evening terminator, therefore overestimating the transit depth in the gaps between the water bands, particularly at 1.65 and 2.25 µm (black dotted line in \cref{Fig:SpectraBest}).
The averaged spectrum formed by combining morning and evening terminators provides a satisfactory fit of the observations in this wavelength range.

While clouds provide larger transit depth in the gaps between the water bands compared to a cloud-free atmosphere (orange line in \cref{Fig:SpectraBest}), and are therefore required to match the observations in the IR, hazes are primordial to provide a good fit of the UV-visible range.
As seen in \cref{Fig:SpectraBest}, the nominal case (black line) provides a steeper UV slope with higher transit depths compared to the haze-free (pink line) and clear atmosphere (yellow line) models.
The nominal model is in agreement with both \cite{Sing16} and \cite{Fischer16} HST observations in the UV, providing residuals within the 3$\sigma$ of the observations, despite the apparent disagreement between these two datasets.
Indeed, \cite{Sing16} analysis of these HST observations indicates larger transit depths and a slightly steeper UV slope than \cite{Fischer16} analysis.
Our nominal model however is consistent with both due to their relatively large error-bars.
We observe in the near-infrared, up to 2 µm, that both haze-free (pink line) and cloud-free (orange line) models provide lower transit depths in the gaps between the water bands compared to the nominal spectrum (black line), indicating that both haze and clouds affect this region.
This is confirmed with the clear model (yellow line), which includes neither haze nor cloud opacities, and provides even smaller transit depths in the gaps between the water bands compared to both haze-free and cloud-free models.
On the other hand, the UV-visible range is dominated by haze opacities appearing higher in the atmosphere (p < 1mbar) compared to cloud opacities.
Clouds therefore have little impact in this region of the spectrum and the cloud-free (orange line) model demonstrates small deviations in this range compared to the nominal case (black line).
We however note that clouds affect the visible region between Na and K lines.
This is related to lower haze opacities in this range compared to the UV, allowing to probe the cloud top.
We further note that the clouds have a stronger effect on the UV range for a haze-free atmosphere related to the absence of haze opacities hiding the effects of the clouds.
Indeed, while the differences observed when including haze opacities (between the cloud-free: orange line, and the best-fit: black line), are negligible, the modifications of the UV wavelength range brought by cloud opacities, without haze extinction, is much stronger (between the clear: yellow line, and the haze-free: pink line).

\subsection{Gas phase constraints}
\label{Sec:Chemistry}

The formation and rainout of the condensates have major ramifications for the chemical composition of the atmosphere especially the species depleted by the formation of these clouds.
In the saturated region of the atmosphere, the condensing species are consumed to cloud formation until the species partial pressure is roughly equal to its saturation pressure.
The formed particles settle and reach the undersaturated region below where they sublimate, thus releasing the material previously accreted.
This material can then be lifted up by transport via mixing to refill the saturated region in condensing material.
This process reaches a steady state when the upward flux of condensing material compensates for the downward flux of particles settling and transport has quenched the atmosphere above the saturated region to the saturation abundance.
We therefore observe lower Na and H$_2$S (SiO and H$_2$O) mixing ratios at and above the Na$_2$S (MgSiO$_3$) cloud base relative to a cloud-free atmosphere (\cref{Fig:Best}). 
We note that the species used as proxy (Na for Na$_2$S and SiO for MgSiO$_3$) are strongly depleted and only traces of these species remain in the upper atmosphere.
On the other hand, H$_2$O and H$_2$S are too abundant to be strongly impacted by the cloud formation and remain present in large amounts.
We note a drop of 25\% for water, and negligible for H$_2$S (\cref{Fig:Best}).

The chemical composition is directly affected by the change of temperature between the morning and evening terminators due to modifications in the rates of the different reactions.
The same apply to the saturation pressure for the species serving as proxy for cloud formation.
This is well observed for sodium whose equilibrium partial pressure on the evening side is lower than the saturation pressure, preventing its condensation.
We therefore obtain 10$\times$solar abundance of sodium, with a Na mixing ratio of 3$\times$10$^{-5}$ above the 100 bar altitude (\cref{Fig:Best}), for the evening terminator, while on the morning side, the sodium has been depleted by the formation of Na$_2$S condensates, thus providing a Na mixing ratio lower than 10$^{-8}$ for the nominal case above the Na$_2$S cloud formation altitude at 10 mbar (\cref{Fig:Best}).
This four orders of magnitude lower Na mixing ratio for the morning case results in a weak Na line underestimating the transit depth, while the unmuted Na feature on the evening terminator overestimates the transit depth (\cref{Fig:SpectraBestNa}).
The averaged spectrum of the two terminators results in a weaker and narrower Na line (best-fit: black line in \cref{Fig:SpectraBestNa}) compared to the model excluding Na$_2$S formation (Na$_2$S-free model: blue line in \cref{Fig:SpectraBestNa}), therefore resulting in a good fit of the observations and solving for the discrepancies previously observed.
In the formation of Na$_2$S condensates, sodium is the limiting species and H$_2$S abundance is therefore weakly impacted by the modifications of the cloud formation. 
The variations observed for H$_2$S above 1 mbar in \cref{Fig:Best} between the two terminator cases are more likely related to the differences of temperature structure and reflect the changes in the chemical reaction rates for that species.
SiO is not strongly affected by the change of temperature profile, though we note a difference in its abundance profile above 1 bar between the terminators.
This reflects the evolution of its saturation pressure with altitude in this region.
H$_2$O follows the behavior of SiO, therefore the negligible variations observed for SiO result in weak modifications of the water profile between the morning and evening terminators.
The partial depletion of H$_2$O owing to MgSiO$_3$ formation slightly reduces the strength of the water bands relative to a cloud-free atmosphere, impacting the pressure referencing and therefore the whole spectrum, notably the UV-visible slope.
However, these variations remain small and the impact on the transit depth is negligible (best-fit: black line and MgSiO$_3$-free: green line in \cref{Fig:SpectraBest}).

\begin{figure}
\includegraphics[width=0.5\textwidth]{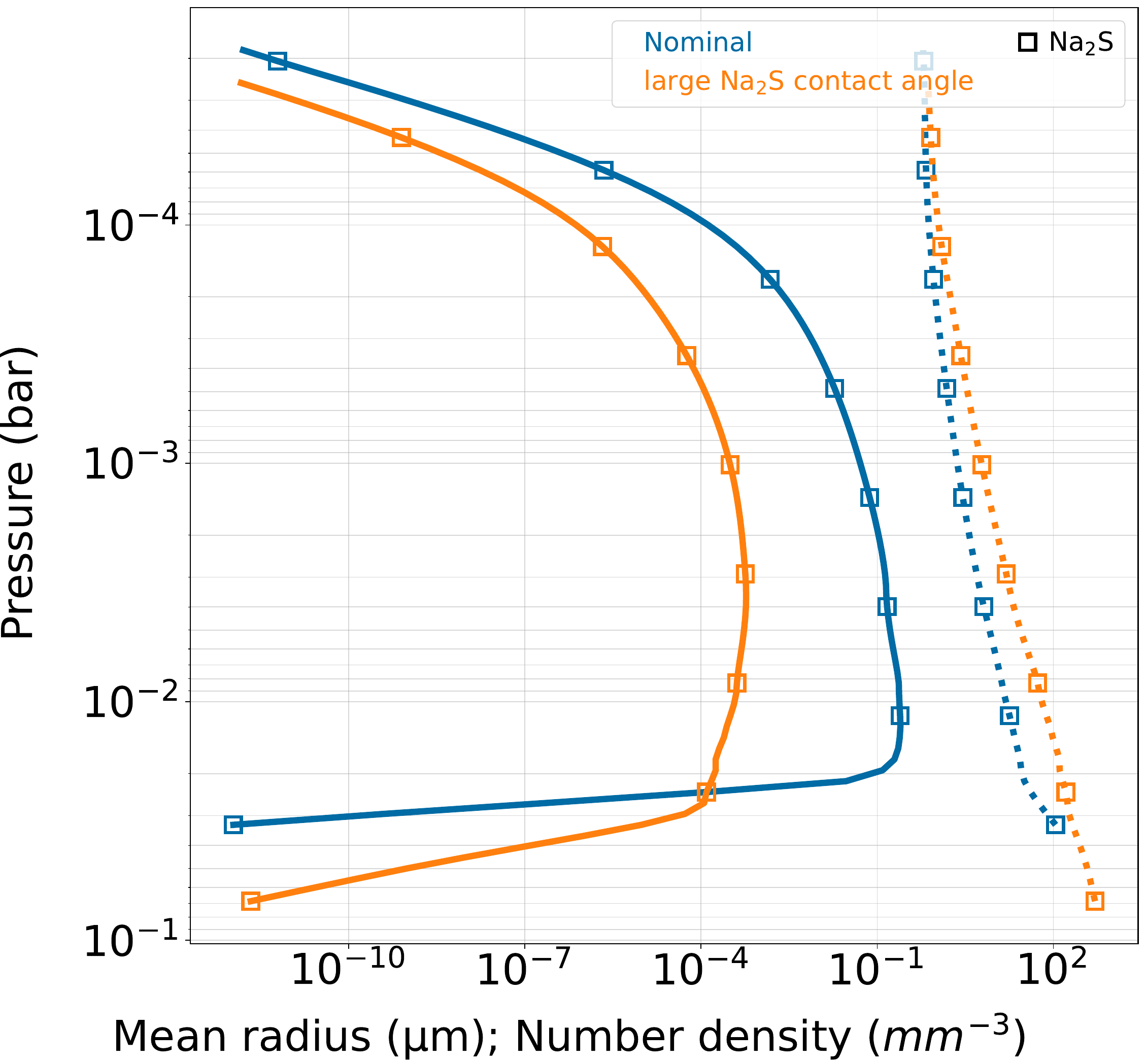}
\caption{Effect of the contact angle for Na$_2$S condensates in the best-fit morning case.
Blue lines are for the low contact angle ($\theta_c$ = 5.7°) and the orange for the large contact angle ($\theta_c$ = 61°).
}
\label{Fig:ContactAngle}
\end{figure}

\begin{figure*}
\includegraphics[width=\textwidth]{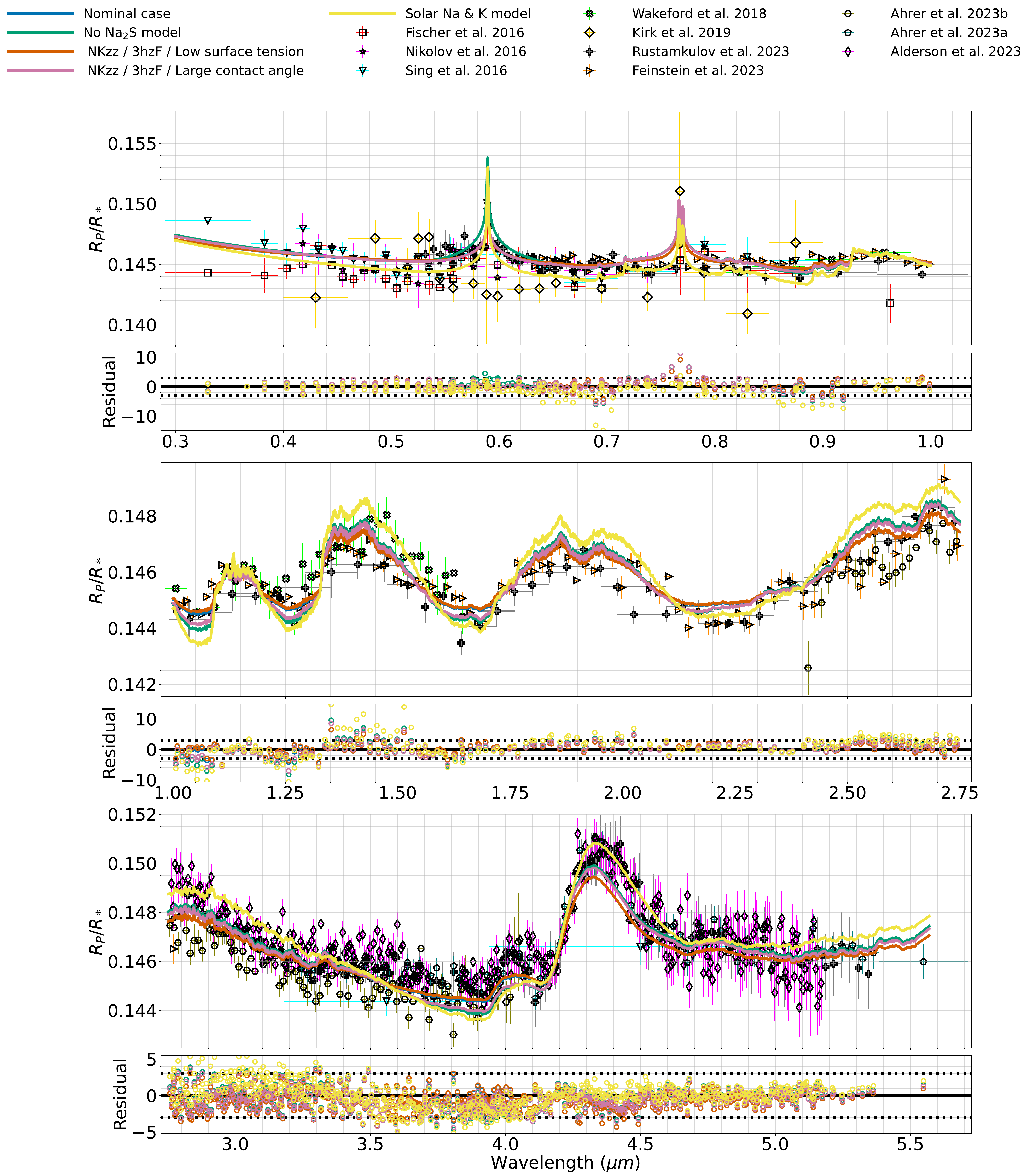}
\caption{Transit spectrum for the best-fit case (blue line), as well as tests with a lower MgSiO$_3$ surface tension (orange line), a larger Na$_2$S contact angle (pink line) and no Na$_2$S formation (green line). 
The yellow line presents the spectrum for a 10$\times$solar metallicity cloud-free model with solar Na and K abundances.
}
\label{Fig:NaK}
\label{Fig:STCA}
\end{figure*}

\begin{figure}
\includegraphics[width=0.5\textwidth]{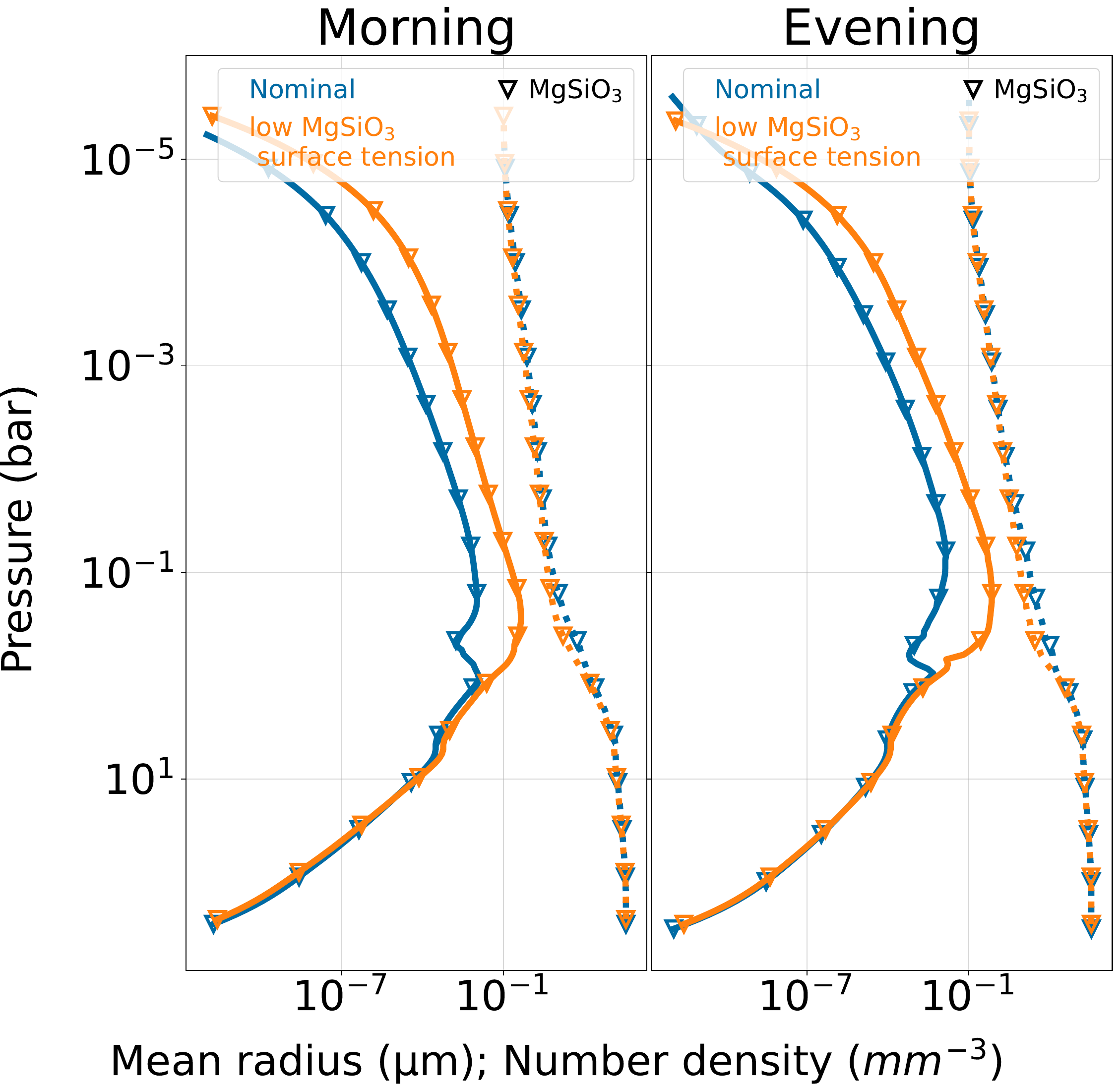}
\caption{Effect of the surface tension for MgSiO$_3$ condensates in the best-fit case.
Solid lines are for the large surface tension ($\sigma$ = 1280 $dyne.cm^{-1}$) and the dotted for the low surface tension ($\sigma$ = 72 $dyne.cm^{-1}$).
The left panel is for the morning terminator and the right panel for the evening.}
\label{Fig:SurfaceTension}
\end{figure}


CO absorption is observable around 4.8 µm. 
In this range, our model provides a good fit of the JWST/NIRSpec PRISM \citep{Rustamkulov23,Ahrer23a} and JWST/NIRSpec G395H \citep{Alderson23} observations, indicating that CO is present in super-solar abundance.
Our model provides a CO mixing ratio of 4$\times$10$^{-3}$, similar to the value retrieved by \cite{Grant23}.

CH$_4$ has not been detected in WASP-39b atmosphere, which is in agreement with our results.
Indeed, our model indicates that methane is photo-dissociated at 1 mbar altitude for the evening case and 0.1 mbar for the morning (\cref{Fig:Best}).
Therefore, our theoretical transit spectra do not demonstrate the CH$_4$ absorption band at 3.3 µm and provide a good fit to the observations.

In the water band, from 0.9 to 3.5 µm, our best-fit spectrum is mostly within the 3$\sigma$ of the observations from HST/WFC3 \citep{Wakeford18}, JWST/NIRSpec PRISM \citep{Rustamkulov23,Ahrer23a}, JWST/NIRCam \citep{Ahrer23b} and JWST/NIRSpec G395H \citep{Alderson23}.
However, we slightly overestimate the JWST/NIRISS observations \citep{Feinstein23} in the water bands (especially around 1.4 µm) while we underestimate them in the gaps between the water bands (around 1.05 and 1.25 µm), which indicates larger cloud opacities.
We derive for the water abundance a super-solar mixing ratio of 2$\times$10$^{-3}$, which is roughly in agreement with the \cite{Fisher18} and \cite{Min20} works, though much larger than the value suggested by \cite{Tsiaras18} and \cite{Pinhas19}.

Our best fit slightly matches the CO$_2$ and SO$_2$ bands compared to observations led by the JWST/NIRSpec PRISM \citep{Rustamkulov23}, JWST/NIRSpec G395H \citep{Alderson23} and JWST/NIRCam \cite{Ahrer23b}, producing residuals within the 3$\sigma$ limit of these observations (\cref{Fig:SpectraBest}).
We however note residuals off the 3$\sigma$ for the JWST/NIRSpec PRISM observations led by \cite{Ahrer23a} near the SO$_2$ line.
We found a maximum mixing ratio of 5$\times$10$^{-5}$ for SO$_2$ on the morning terminator (\cref{Fig:Best}), which is roughly in agreement with the findings of \cite{Alderson23} and \cite{Rustamkulov23}.
The strength and location of this maximum, near 100 µbar, also agrees with the peak in the SO$_2$ profiles derived by \cite{Tsai23}.
Our model indicates that a slightly larger abundance is required to fit the observations, though we note that changes in the referencing, notably related to cloud opacities, can modulate the strength of this feature.
A larger reference pressure can shift the transit spectrum to higher transit depths, thus increasing the strength of the SO$_2$ line relative to the wavelength range used for the referencing.
We also note that decreasing the C/O ratio can increase the SO$_2$ abundances \citep{Tsai23} and improving the fit.
For CO$_2$, we find a mixing ratio of $\sim$10$^{-5}$ at 1 mbar in agreement with the profile derived by \cite{Tsai23} and \cite{Carone23}, who used C/O = 0.55, while \cite{Grant23}, who used a sub-solar C/O ratio of 0.3, found a CO$_2$ abundance of 7$\times$10$^{-6}$, slightly lower than the value we derived.

Our nominal spectrum also provides a good fit with most observations in the visible wavelength range \citep{Ricci15,Fischer16,Sing16,Nikolov16,Kirk19,Rustamkulov23}, except for \cite{Feinstein23} for which the residual peaks at 10$\sigma$ for the potassium line data.
This is not observed for the other datasets probing the K line owing to either large uncertainties or to a broader wavelength integration.
However, the resolution and sensitivity of the JWST observations allow a more precise measurement of the potassium feature and indicate lower mixing ratios than the 2$\times$10$^{-6}$ provided by the 10$\times$solar metallicity used in our simulations.
We discuss this issue further below.

\section{Sensitivity tests}
\label{Sec:Sensitivity}

In the following of this study, we aim to describe how changes in the surface tension, contact angle, eddy diffusion and haze production impacts the formation of the clouds, as well as the ramifications for the chemistry and transit spectra.



\subsection{Impact of cloud properties}
\label{Sec:STCA}

As discussed in \cref{Sec:CldCompo}, the contact angle and surface tension of the condensates are not constrained and we aim here to evaluate their impact on cloud formation.

We made a test with the contact angle value of 61° derived by \cite{Gao20} for the nucleation of Na$_2$S on TiO$_2$.
This provides much weaker nucleation rates resulting in a particle number density up to 3 orders of magnitude smaller compared to the results obtain with $\theta_c$=5.7° (\cref{Fig:ContactAngle}).
As a consequence, we get lower transit depths in the gaps between the water bands, providing similar results as a Na$_2$S-free atmosphere (\cref{Fig:STCA}).
On the other hand, the particles formed are larger and the depletion of sodium from the gas phase remains strong.
Therefore, the large Na$_2$S contact angle case keeps providing a good fit of the Na line (\cref{Fig:STCA}).
Obtaining constrains on the contact angle for the formation of condensates over haze particles is therefore primordial to support the hypothesis of haze serving as nucleation sites and we stress that lab experiments are required.

The surface tension of the condensates included in this study are not well constrained either and the values we use are rough estimates.
Especially, the value of 1280 $dyne.cm^{-1}$ we used for MgSiO$_3$ is estimated from Mg$_2$SiO$_4$ \citep{Leeuw00} and the use of a different value can affect the results.
For MgSiO$_3$, other sources indicate a much smaller surface tension of $\sim$80 $dyne.cm^{-1}$ \citep{Voelkel94}, while \cite{Gao20} uses a value of 436 $dyne.cm^{-1}$ for Mg$_2$SiO$_4$. 
\cref{Fig:SurfaceTension} presents the cloud distributions obtained with the $\sim$80 $dyne.cm^{-1}$ surface tension compared to the profile obtained with the 1280 $dyne.cm^{-1}$ value.
The small surface tension value enhances the production of the cloud, thus increasing both nucleation and condensation rates leading to more numerous particles.
We note however that the particle size decreases as the condensing material is distributed over a larger number of particles.
On the other hand, as previously discussed, MgSiO$_3$ has negligible effects on the spectrum and the changes brought by the lower surface tension are not sufficient to affect the impact of MgSiO$_3$ on the transit spectrum.
Moreover, these modifications do not affect H$_2$O and have minor effects on the transit spectrum (\cref{Fig:STCA}).

\subsection{Impact of Kzz on the cloud distributions}
\label{Sec:Eddy}

In addition to the nominal profile (denoted NKzz), we test a second eddy profile (denoted case 10Kzz) which uses values 10 times larger than the nominal above the radiative/convective boundary (orange line \cref{Fig:Eddy}).

\begin{figure}
\includegraphics[width=0.48\textwidth]{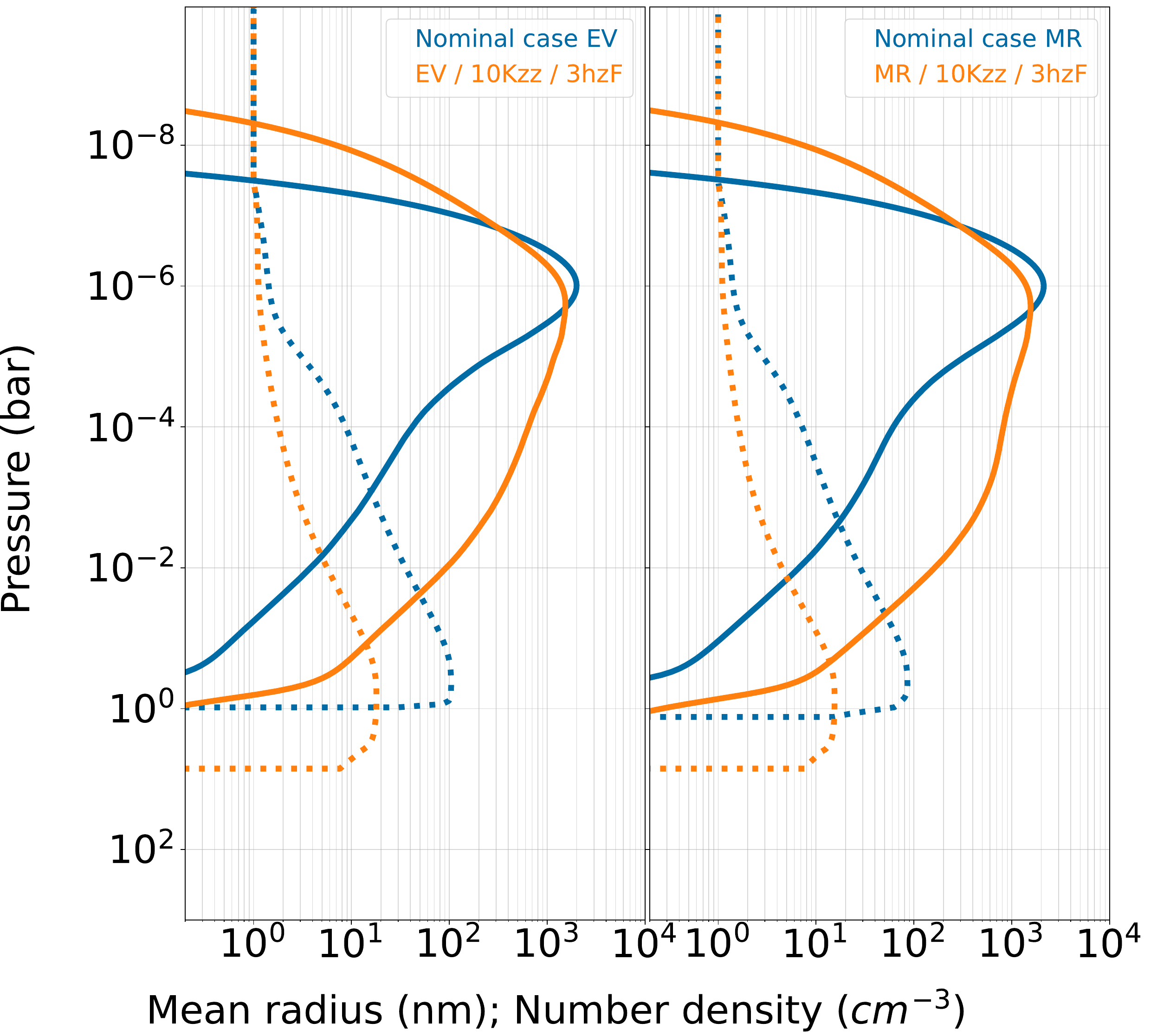}
\caption{
Haze distributions of the different cases tested comparing the nominal (blue lines) and 10$\times$ nominal (orange lines) eddy diffusion cases for the two haze mass flux tested (top: 1hzF, bottom: 3hzF) and for both terminators (left: evening, right: morning).
Solid lines are the number densities and dotted lines the mean particle radii.
}
\label{Fig:HazeKzz}
\end{figure}

\begin{figure}
\includegraphics[width=0.48\textwidth]{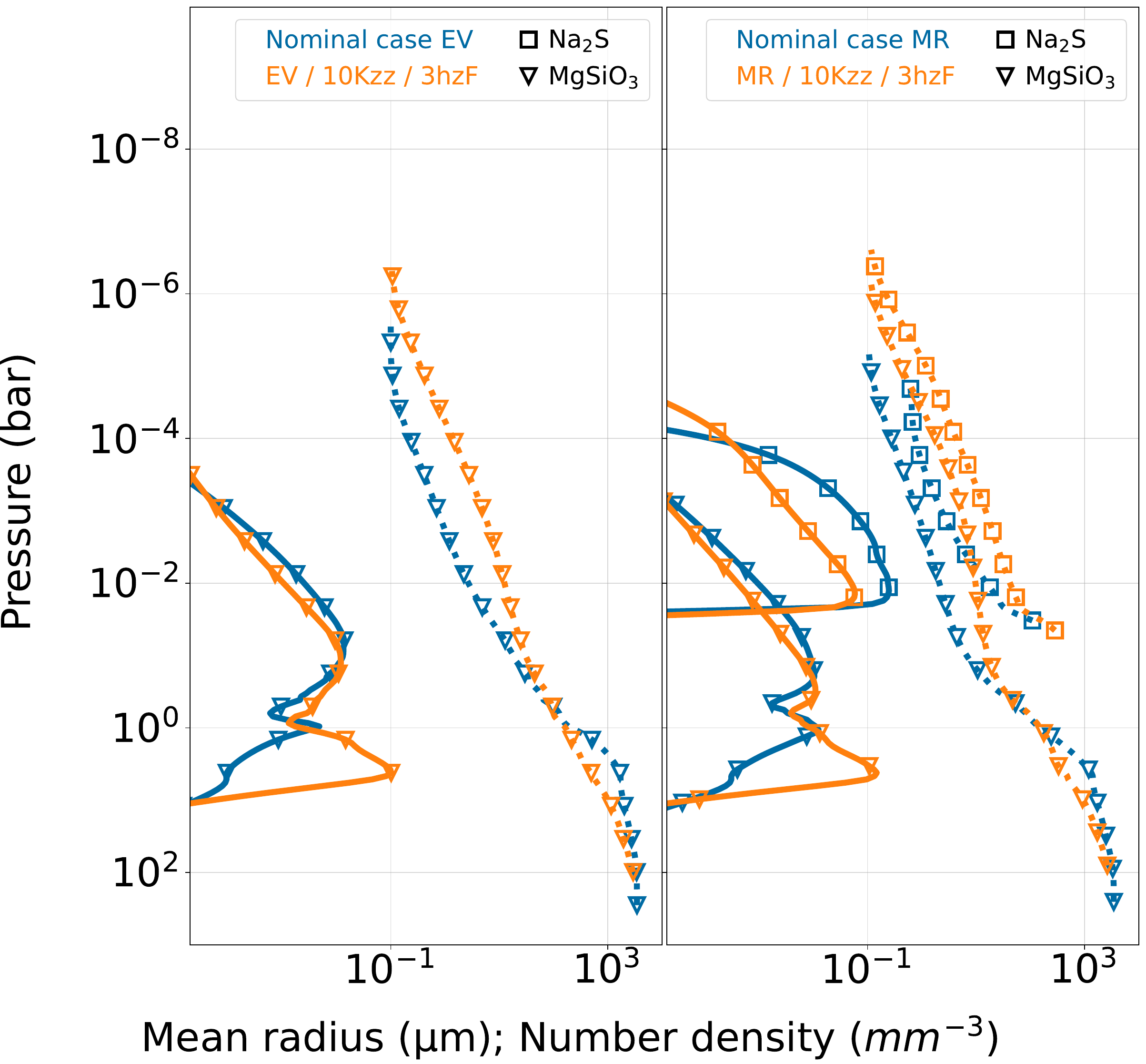}
\caption{Cloud distributions of the different cases tested comparing the nominal (blue lines) and 10$\times$ nominal (orange lines) eddy diffusion cases for the two haze mass flux tested (top: 1hzF, bottom: 3hzF) and for both terminators (left: evening, right: morning).
Solid lines are the number densities and dotted lines the mean particle radii.
Curves with square markers refers to Na$_2$S condensates and triangles to MgSiO$_3$.
}
\label{Fig:CloudKzz}
\end{figure}

\begin{figure}
\includegraphics[width=0.48\textwidth]{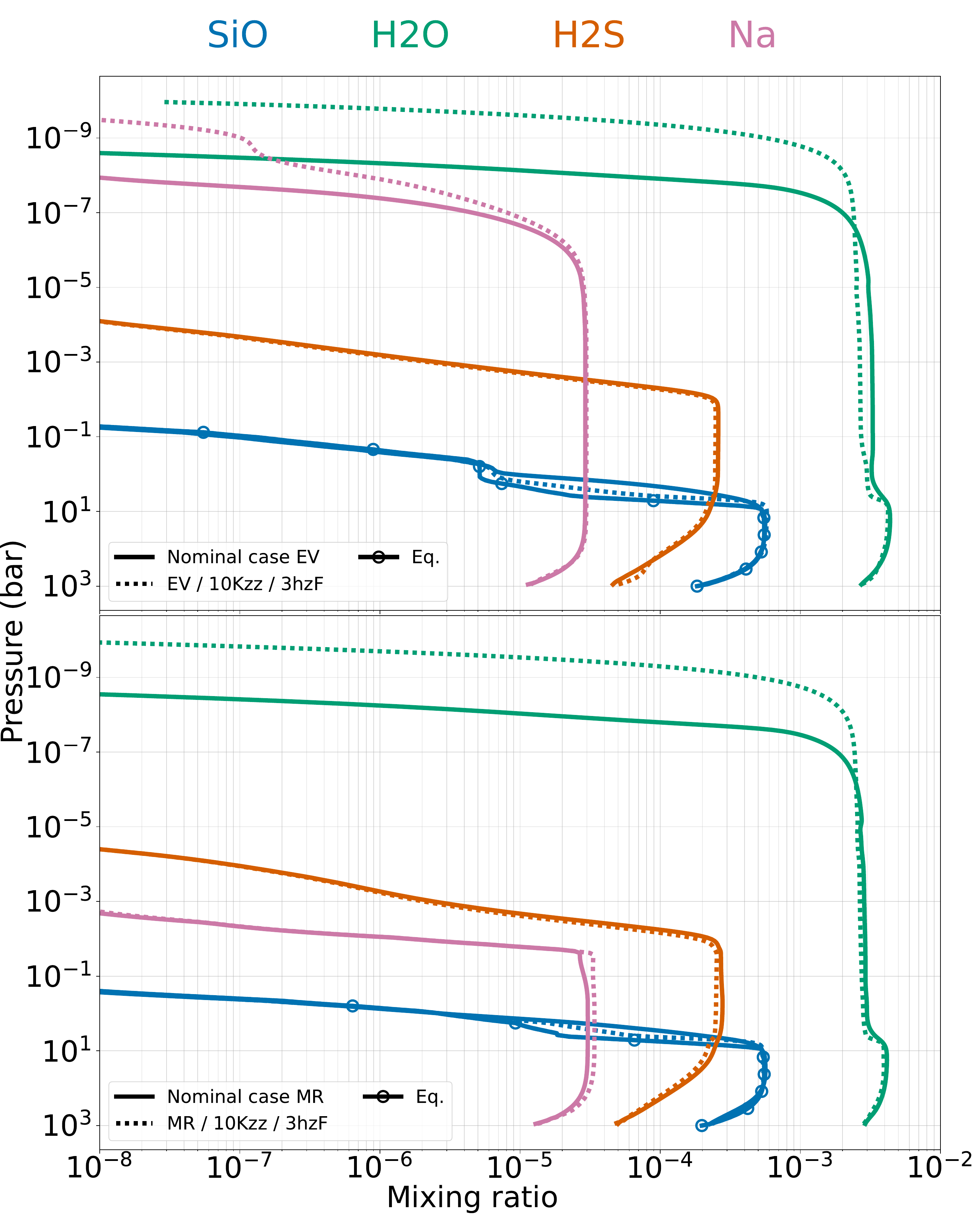}
\caption{Effect of the eddy diffusion profile on the chemical composition profile of the for the 3HzF case, for both terminators (top: evening, bottom: morning).
Circled lines are the thermochemical equilibrium solutions for SiO.
}
\label{Fig:Chemistry1}
\end{figure}


\subsubsection{Na$_2$S}

A larger eddy diffusion results in smaller and more numerous haze particles (\cref{Fig:HazeKzz}) as the transport timescale decreases relatively to the coagulation timescale \citep{Arfaux22}.
These smaller haze particles provide a less efficient nucleation and lead to a smaller Na$_2$S particle density in the 10Kzz cases relative to the NKzz as observed in \cref{Fig:CloudKzz}.
As the condensing material is distributed among less numerous condensates, we observe an increase of the mean particle size.
In addition, the mass flux of condensing species is larger in the 10Kzz case owing to the more efficient mixing by eddy diffusion, resulting in up to 10$\times$ larger condensation rates.
We therefore obtain much larger particles than what would have been produced if the mass flux of sodium was conserved.
The location of the Na$_2$S cloud formation region remains unchanged, however, under the more efficient transport related to the larger eddy diffusion, the clouds expand much higher up in the atmosphere reaching the µbar level.

The formation of Na$_2$S clouds depletes the atmosphere from its sodium content, but very little differences appear between the two eddy cases (\cref{Fig:Chemistry1}), since the mixing ratio reached by the condensing species during cloud formation is limited by their saturation pressure that remains unaffected by the change of eddy diffusion.
The change of eddy therefore leaves the Na composition profile unaffected.
The same conclusion applies for H$_2$S which remains unaffected by the change of eddy.

\begin{figure*}
\includegraphics[angle=-90,width=\textwidth]{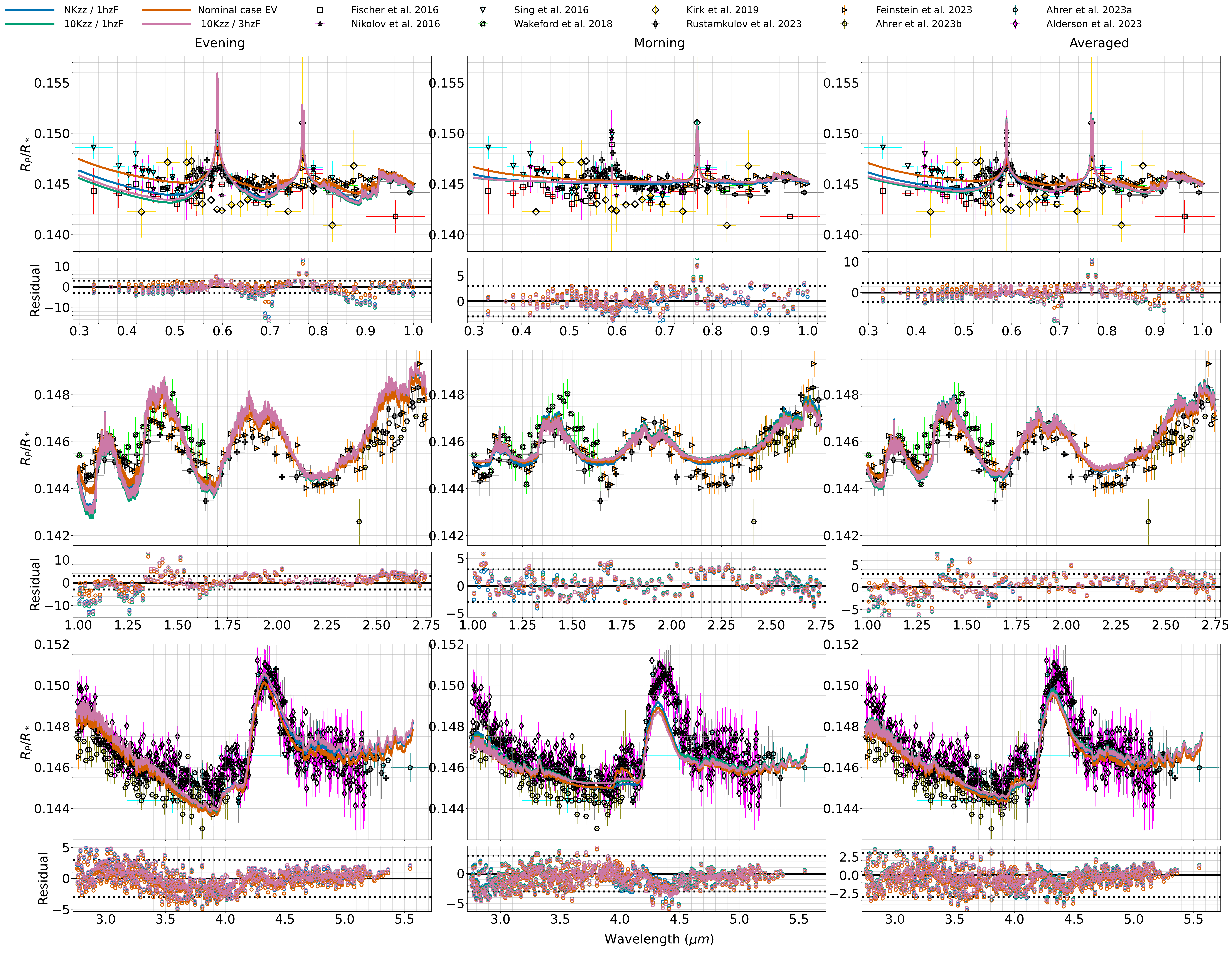}
\caption{Transit spectra obtained for the different cases tested. 
From top to bottom, we present the spectra for the evening cases, morning cases and the averaged.
Overplotted are the different observations available for WASP-39b along with the corresponding residuals.
The spectra are smoothed with a savgol filter.
}
\label{Fig:Spectra}
\end{figure*}

\subsubsection{MgSiO$_3$}

For MgSiO$_3$ condensates, we note a slightly different behavior than Na$_2$S.
While in the NKzz case, the second, deeper, cloud formation region provided only a small peak in the density profile, in the 10Kzz case it becomes the main source of MgSiO$_3$ condensate particles and demonstrates a larger particle number density compared to the above nucleation region (\cref{Fig:CloudKzz}).
This is related to haze particles reaching below the 1 bar altitude in larger abundance compared to the nominal eddy case, enhancing the nucleation rates around 1 bar. 
In addition, this second region is located deeper where the atmospheric density is large and therefore the material for cloud formation abundant, resulting in larger condensation rates in the 10Kzz case compared to the NKzz.
We also observe larger particles above the 0.1 bar altitude in the 10Kzz case compared to the NKzz case related to the stronger eddy which, in one hand enhances the flux of condensable material, and in the other hand, more efficiently lifts large particles.



In the thermochemical solution the formation of MgSiO$_3$ starts in the region of the atmosphere below the "undersaturated" region and the depletion of SiO happens at 10 bar.
However, in our calculations, due to the low abundance of condensation nuclei below the "undersaturated" region, the depletion of SiO is limited and the profiles do not match the thermochemical solution (\cref{Fig:Chemistry1}).
On the other hand, the shift of the maximum number density in the 10Kzz cases, produces a SiO profile closer to the thermochemical equilibrium solution.
We note that TiO$_2$ particles might be expected in the deep atmosphere of this planet \citep{Carone23} and might serve as CCN, therefore allowing such an important formation of MgSiO$_3$ condensates below the "undersaturated" region in the nominal eddy case as well.



%


\subsubsection{Spectra}

As discussed above, the changes brought by the different eddy mixing profiles on the MgSiO$_3$ distribution are negligible in the upper atmosphere, therefore resulting in limited modifications of the spectra related to MgSiO$_3$ cloud opacity (\cref{Fig:Spectra}).
For the morning terminator, the grey opacity provided by Na$_2$S condensates hide any potential modification and we note little variations among the different spectra (\cref{Fig:Spectra}).
Indeed, the reference pressure between these cases remain similar, which indicates that the clouds become optically thick at a similar altitude in both eddy cases.
As a result, the averaged spectrum remains unaffected by the changes in cloud distributions brought by increasing the eddy diffusion.
On the other hand, the changes in the haze distribution brought by the change of eddy diffusion affect the transit spectra, especially in the UV-visible.
The smaller particles related to the larger eddy diffusion result in lower haze opacities in the upper atmosphere.
As a consequence, the region probed by the observations is shifted to larger pressures, therefore producing lower transit depths, and the spectrum is close to a haze-free atmosphere.
The nominal eddy profile is then in better agreement with the HST \cite{Sing16} observations in the UV-visible range compared to the high eddy case.

\subsection{Impact of haze mass flux on the cloud distributions}
\label{Sec:HzMF}



\begin{figure}[t]
\includegraphics[width=0.48\textwidth]{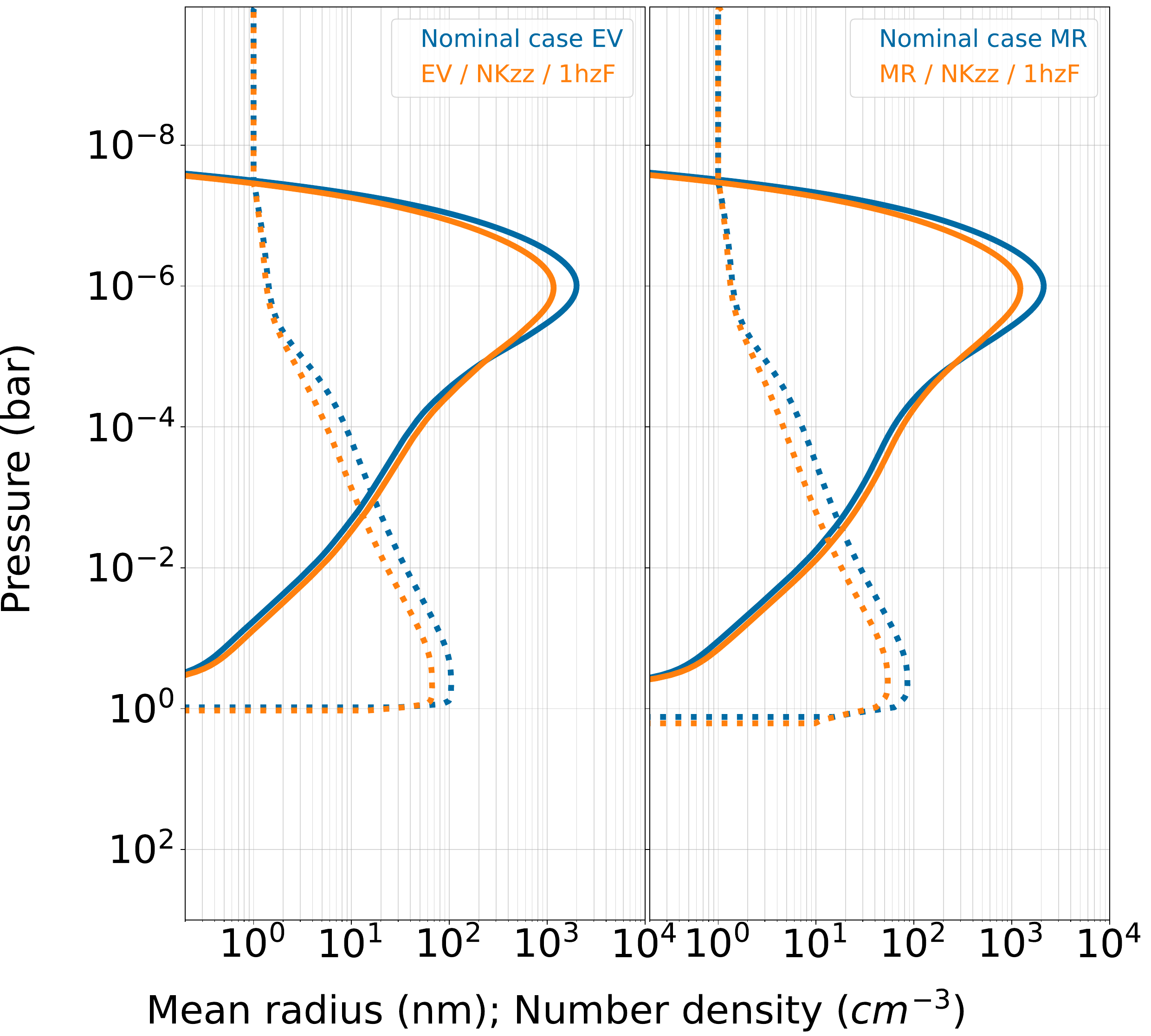}
\caption{Haze distributions of the different cases tested comparing the 3$\times$10$^{-15} g.cm^{-2}.s^{-1}$ (blue lines) and 10$^{-15} g.cm^{-2}.s^{-1}$ (orange lines) haze mass flux cases for the two eddy diffusion profile tested (top: NKzz, bottom: 10Kzz) and for both terminators (left: evening, right: morning).
Solid lines are the number densities and dotted lines the mean particle radii.
}
\label{Fig:HazeHz}
\end{figure}

\begin{figure}[t]
\includegraphics[width=0.48\textwidth]{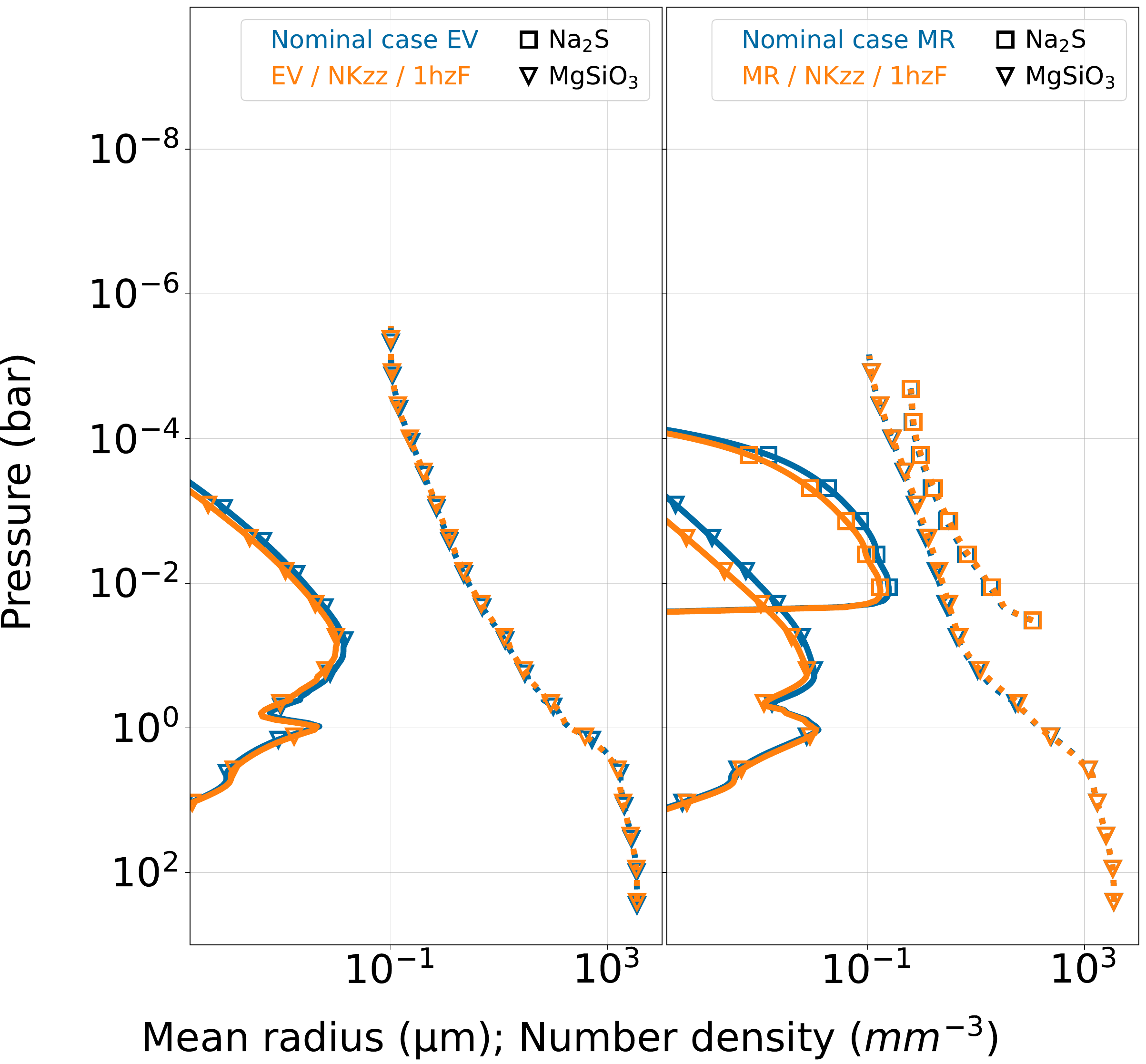}
\caption{Cloud distributions of the different sensitivity cases tested comparing the 3$\times$10$^{-15} g.cm^{-2}.s^{-1}$ (blue lines) and 10$^{-15} g.cm^{-2}.s^{-1}$ (orange lines) haze mass flux cases for the two eddy diffusion profile tested (top: NKzz, bottom: 10Kzz) and for both terminators (left: evening, right: morning).
Solid lines are the number densities and dotted lines the mean particle radii.
Curves with square markers refers to Na$_2$S condensates and triangles to MgSiO$_3$.
}
\label{Fig:CloudHz}
\end{figure}

We tested two different haze mass fluxes: 10$^{-15} $ (case 1HzF) and 3$\times$10$^{-15} g.cm^{-2}.s^{-1}$ (case 3HzF).
Decreasing the haze mass flux results in less numerous particles in the haze formation region (around 1 µbar) but weaker coagulation rates deeper down and therefore smaller particles below the haze production region (\cref{Fig:HazeHz}).
We therefore obtain smaller haze particles in the cloud formation region, with a similar number density, thus hampering the nucleation rates and producing less numerous cloud particles.
However, the effect is rather faint for MgSiO$_3$  (\cref{Fig:CloudHz}) where the 30\% weaker nucleation rates result in $\sim$30\% less numerous MgSiO$_3$ condensate particles, for both terminators.
Na$_2$S formation is also impacted by this change of haze mass flux, with variations of the number density of $\sim$30\%.

These small variations in the cloud distribution under different haze mass fluxes produces negligible variations of the cloud opacity and therefore do not affect the spectra (\cref{Fig:Spectra}).
However, the differences in the haze distribution can have an impact.
For the NKzz case, the difference in particle size affects the UV region, resulting in a slightly steeper slope in the 3hzF case (orange line in \cref{Fig:Spectra}) relative to the 1HzF case (blue line in \cref{Fig:Spectra}) for the both terminators.
On the evening side, we note much larger transit depths for the 3hzF compared to the 1HzF case in the UV-visible, as well as, in the gaps between the water bands around 1.05 and 1.25 µm.
On the other hand, for the 10Kzz cases, changing the haze mass flux has a relatively weak impact on the spectra (green and pink lines in \cref{Fig:Spectra}).
This is due to the lower haze opacity observed in the high eddy cases, resulting in a spectrum close to a haze-free atmosphere.


%
%

%

\section{Discussion}
\label{Sec:Discussion}

\subsection{Clouds interaction and thermal effects}

Our radiative transfer simulations do not include the effect of clouds as they are not anticipated to sufficiently affect the UV part of the radiation field to impact the chemical distribution.
We effectively observe that below 10 mbar, the amount of UV radiation received does not significantly impact the chemical composition of the atmosphere.
However, the presence of clouds at these altitudes could affect the thermal structure of the planet.
In our simulations, the p-T profiles of the terminators have been kept fixed based on the GCM results of \cite{Tsai23}, though the presence of haze and clouds could have an impact on the temperature \citep{Marley13,Lavvas21}.
3D GCM simulations are required in order to include the haze and cloud feedbacks on the thermal structure, accounting for horizontal gradients in their composition and distribution \citep{Lee16,Steinrueck21,Komacek22a}.
Haze absorbs in the UV in the upper atmosphere and heat up that region of the atmosphere \citep{Lavvas21,Arfaux22} while clouds, deeper down, may absorb in the visible and infrared, thus locally increasing the temperature as well. 
This increase of the temperature may affect the cloud formation and distribution and clouds radiative feedback have to be self-consistently accounted.
We note that the secondary MgSiO$_3$ layer observed in our results arise from temperature profile effects.
A lower temperature may allow nucleation in the currently "undersaturated" region and connect the two formation region as one, while a hotter temperature, as may 	arise including haze and cloud feedbacks, can hamper the apparition of this second region.


Our model excludes the interactions between MgSiO$_3$ and Na$_2$S.
Indeed, we can expect MgSiO$_3$ condensates to form on Na$_2$S nuclei.
However, considering the differences in formation altitude, this interaction is not expected to play a major role as Na$_2$S will dominate the mixed clouds in the upper atmosphere and MgSiO$_3$ will dominate in the deep atmosphere.

We note that multiple studies consider the possibility of MnS acting as part of the missing opacities in WASP-39b's atmosphere \citep{Ahrer23a,Ahrer23b,Alderson23,Feinstein23}.
Based on preliminary results using ggChem on the morning terminator, this cloud species forms at larger pressures (1 bar) compared to Na$_2$S (10 mbar) and remains less abundant than Na$_2$S by at least 6 orders of magnitude.
MnS is therefore not expected to affect the morning terminator.
However, MnS forms higher than MgSiO$_3$ (0.1 bar against 1 bar for MgSiO$_3$) on the evening terminator and expands to the mbar level (against 10 mbar for MgSiO$_3$).
MnS clouds could therefore affect this terminator.
We note that, for both terminators, MnS is the dominating cloud species from 10 to 100 mbar, which is below the pressure range probed by the observations.
Therefore, MnS may have locally a higher opacity than MgSiO$_3$, but  but Na$_2$S should still dominate the cloud opacities.
Therefore, the inclusion of MnS is not expected to present major ramifications for the transit spectrum.


\subsection{Other hypothesis}

Other hypothesis may explain the observations and we aim to discuss them in this section.

%


%

\begin{figure}
\includegraphics[width=0.5\textwidth]{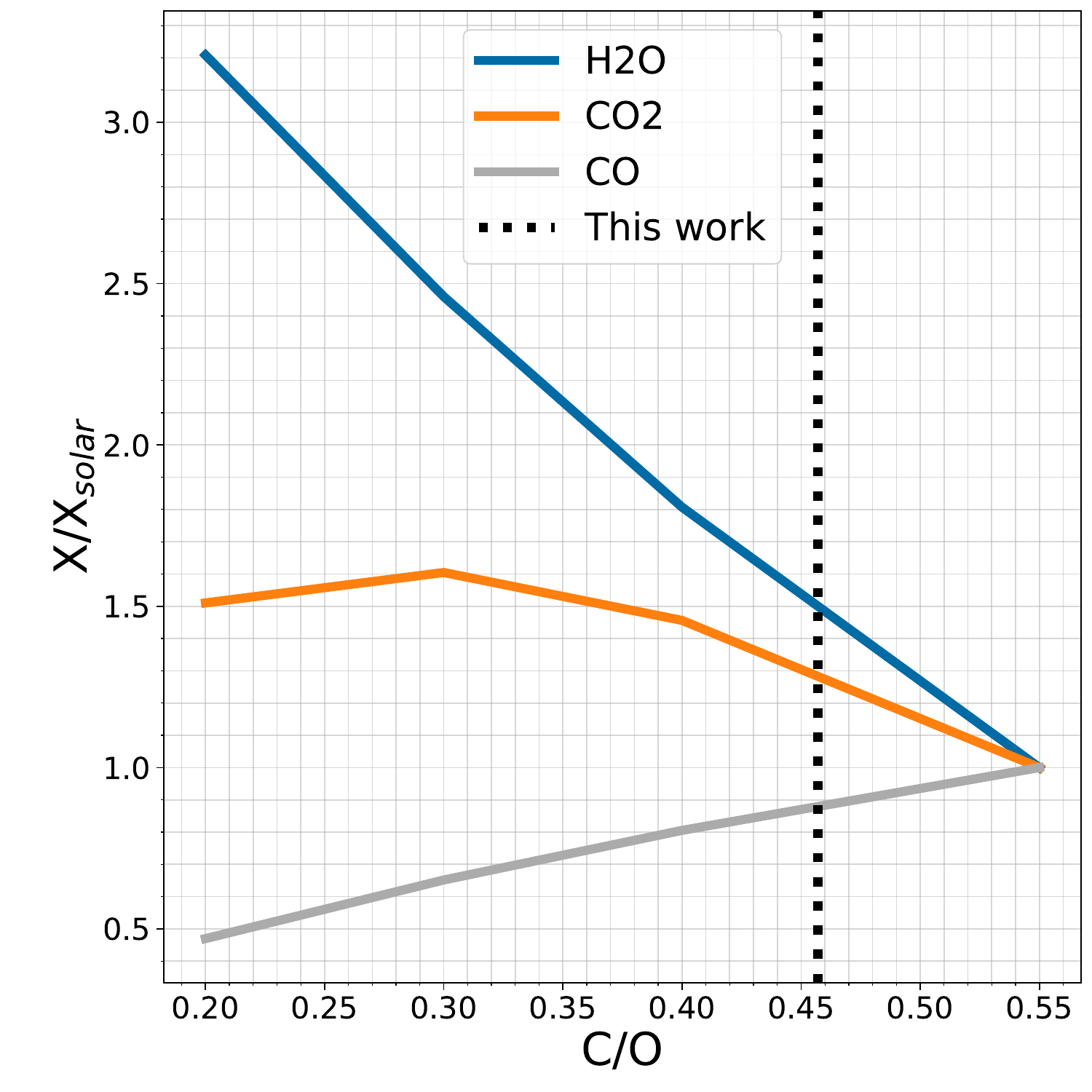}
\caption{Thermochemical equilibrium mixing ratios of H$_2$O (blue lines), CO (grey lines) and CO$_2$ (orange lines) at different assumed values of C/O abundance ratio relative to the value obtained with a solar C/O.
The values are taken at the quench level of the different species (0.02 bar for H$_2$O, 0.1 bar for CO and 0.2 bar for CO$_2$) in the nominal case for the evening terminator conditions.}
\label{Fig:CO}
\end{figure}

The recent JWST observations, as well as, some previous works indicate a C/O ratio ranging from 0.2 to 0.55 \citep{Wakeford18,Kawashima21,Ahrer23a,Ahrer23b,Alderson23,Feinstein23,Crossfield23,Grant23}.
In this study, we used a value of of 0.457, but we note that lower values can be expected and could help in improving the fit of the CO$_2$ band.
Simulations with the GGchem model demonstrate a 25\% larger CO$_2$ abundance for a C/O ratio of 0.3, relative to our simulated CO$_2$ abundance for a C/O=0.457 (at the CO$_2$ quench level \cref{Fig:CO}).
While our best-fit spectrum slightly underestimates the strength of the CO$_2$ line (\cref{Fig:SpectraBest}), using a smaller C/O ratio may therefore help to produce a better fit of the CO$_2$ absorption band.

Our best-fit spectrum largely overestimates the strength of the K line (\cref{Fig:SpectraBest}).
A hypothesis for this low potassium abundance could be a primary depletion during planet formation resulting in solar or sub-solar potassium densities.
However, our results indicate that a 10$\times$solar sodium abundance provides a good fit of the spectrum and we expect potassium to undergo a similar enrichment as Na, based on observations of other astronomical objects \citep{Lavvas14}.
We note the results from \cite{Pinhas18} who found degeneracies between the Na and K abundances and the stellar contamination.
They demonstrate that stellar variability can have an impact on the retrieval of the atmospheric alkali composition, especially cold spots can mimic the Na and K features.
However, WASP-39b transit spectrum is assumed free from stellar contamination \citep{Faedi11,Sing16,Fischer16,Ahrer23b,Rustamkulov23}.
Another hypothesis is the loss of potassium to cloud formation as observed for sodium in the current study.
However, potassium does not form any condensate in the pressure/temperature conditions of WASP-39b.
Adsorption of potassium into other condensates may partly deplete the atmosphere from its potassium content and improve the fit of the K line.
This process is however poorly studied and no constraints are available to test for this hypothesis. 
Finally, a possibility is a primary depletion of both alkali elements during the planet formation.
Considering the low metallicity of the host star \citep[-0.12 dex, ][]{Faedi11}, weak (solar) abundances of alkali elements are a possibility despite the enrichment expected for the other elements.
A preliminary test using a solar metallicity for the alkali elements while keeping the 10$\times$solar metallicity for the other species, is consistent with the observations within the 3$\sigma$ for the Na line but surpasses the 4$\sigma$ limit for the K line (\cref{Fig:NaK}).
This test uses a cloud-free self-consistent model, including haze feedback on the temperature profile and assuming a full heat redistribution \citep[Section 2.4 in][]{Arfaux23}.
We therefore note that the disagreement with the potassium feature is not solved and that clouds are still required in the water bands and for the continuum between the Na and K lines to explain the observed transit.

\subsection{Effects of terminator differentiation on the haze precursors}

The difference in temperature between the terminators can have an impact on the chemical species considered as precursors for the photochemical haze formation.
Particularly, we consider HCN to be the main haze precursor for hot-Jupiters with equilibrium temperature lower than 1300 K \citep{Arfaux22}.
In \cref{Fig:Best}, we observe that the morning terminator demonstrates larger HCN abundances related to the lower temperatures.
Indeed, HCN formation arises from the photolysis of NH$_3$, which presents larger abundances as the temperature decreases, owing to a larger quenching mixing ratio \citep{Arfaux23}.
As a result, the morning terminator presents HCN abundances larger by up to 2 orders of magnitude and provides a photolysis mass flux of HCN 100 times larger, compared to the evening terminator.
Therefore, it is likely to observe a larger abundance of haze on the morning terminator, while we assume the same haze production in both terminators.
However, we must keep in mind that the temperature has an unknown effect on the processes leading to the formation of the haze particles, and therefore different haze formation yield can apply between the terminators.
Also dynamics affect the distribution of the haze particles \citep{Steinrueck21,Steinrueck23}. 
Such aspects require coupling of the microphysics with a 3D GCM. 

%

\section{Conclusions}
\label{Sec:Conclusions}

Our results indicate that both haze and clouds are required to produce a satisfactory fit of the HST and JWST observations. 
Under the physical properties (surface tension, contact angle, etc.) assumed for the nucleation of Na$_2$S and MgSiO$_3$ over haze particles, we demonstrate that the nucleation over haze particles leads to a significant production of clouds with detectable effects on the spectrum.
The formation of MgSiO$_3$ condensates has negligible effects on the spectra, despite the removal of water affecting the transit spectra, and Na$_2$S therefore dominates the cloud opacities.
Na$_2$S cloud opacity demonstrate a gray absorber behavior matching the water bands observed by both HST and JWST.
We also note that the removal of sodium provides a good fit of the Na line observations.
We further highlight the need to consider both terminators since a complete depletion of sodium would have underestimated the Na line, as demonstrated by the morning terminator spectra, while the Na contribution to the spectrum from the evening terminator is still required.
Our results provide a best-fit with a haze mass flux of 3$\times$10$^{-15} g.cm^{-2}.s^{-1}$ and the nominal eddy profile.
This best fit is in agreement with most WASP-39b transit observations.
Our simulations are in agreement (within 3$\sigma$) with the presence of sulfur and carbon dioxides, though we note that our best-fit slightly underestimates the SO$_2$ and CO$_2$ signatures. 
The agreement may be further improved by modifying the C/O ratio.
The model also indicates that methane is lost due to photolysis reactions, which agrees with the non-detection of this species in the observed transit spectrum.
Finally, the potassium abundances produced by our model lead to an overestimation of the transit depth at the K line.
A depletion of potassium via adsorption in cloud particles formed by another condensate is a plausible explanation, though further theoretical and experimental studies are required to test this hypothesis.

Our nominal case uses a large surface tension of 1280 $dyne.cm^{-1}$ for MgSiO$_3$ and a small contact angle of 5.7° for Na$_2$S.
Sensitivity tests with lower surface tension for MgSiO$_3$ condensates indicate that the evening terminator is strongly affected with much weaker nucleation rates providing fewer but larger particles.
The morning terminator is however not strongly affected by the change of MgSiO$_3$ surface tension.
An additional test with a larger contact angle of 61° for Na$_2$S results in a significant drop of this condensate density, resulting in a transit spectrum close to a cloud-free. 
We however note that the depletion of sodium to cloud formation remains strong and the residuals at the Na line remain within the 3$\sigma$ of the observations.

We further demonstrate that modifications of the eddy profile can have major ramifications for the cloud formation.
For MgSiO$_3$ condensates in WASP-39b atmosphere we observe that the second formation region becomes more important, presenting larger particle number number densities, in relation to the larger haze abundance due the stronger eddy diffusion.
However, this is explained by a "undersaturated" region whose presence strongly depends on the temperature structure.
For Na$_2$S, the smaller haze particles produced by the stronger eddy diffusion result in a smaller cloud number density, though larger cloud particles compared to the nominal eddy case.
However, these variations of the cloud distributions related to changes in the eddy diffusion have negligible effects on the transit spectra.

Changing the haze mass flux has little impact on the cloud formation. 
The decrease of the haze mass flux by a factor of 3 results in nucleation rates only $\sim$30\% weaker.
This can be understood by the decrease of haze particle size instead of number density, resulting in weaker effects in the nucleation process.
We however note that using a larger haze mass flux has ramifications for the spectra, not only in the UV-visible range where it produces larger transit depths in agreement with HST and VLT observations, but in the water band as well since haze absorption affects the NIR part of the spectrum.

We highlight that knowledge about haze and clouds physical properties (like surface tension) is required to draw more precise results and we stress that lab experiments on the nucleation of cloud species on soot-type aerosols are required to set definitive constraints on the haze and clouds coupling in hot-Jupiter atmospheres.
We further note that 3D modeling accounting for haze and cloud radiative feedback might result in modifications of the cloud distributions and can provide further insights on the clouds properties in WASP-39b atmosphere.


\section*{Acknowledgements}

\section*{Data Availability}

 The data underlying this article will be shared on reasonable request to the corresponding author.



\bibliographystyle{mnras}
\bibliography{biblio} 




%


\bsp	
\label{lastpage}
\end{document}